\documentclass[12pt]{article}

\usepackage{amsfonts}
\usepackage{pslatex}

\makeatletter

\@addtoreset{equation}{section}
\makeatother

\def\be{\begin{equation}}
\def\ee{\end{equation}}
\def\bea{\begin{eqnarray}}
\def\eea{\end{eqnarray}}

\def\({\left(}
\def\){\right)}
\def\<{\left<}
\def\>{\right>}

\def\tr{{\mbox{tr}}}
\def\be{\begin{equation}}
\def\ee{\end{equation}}
\def\bea{\begin{eqnarray}}
\def\eea{\end{eqnarray}}
\def\ben{\begin{eqnarray}}
\def\een{\end{eqnarray}}
\def\({\left(}
\def\){\right)}
\def\<{\left<}
\def\>{\right>}

\def\[{\left[}
\def\]{\right]}

\def\+{\bar}
\def\mb{\mathbb}
\def\tr{{\mbox{tr}}}

\def\t{\widetilde}

\def\N{{\cal{N}}}

\def\rmd{{\rm d}}

\begin{document}

\pagestyle{empty}
\vskip-3cm
\rightline{SNUST-090601}
\vskip0.5cm
\begin{center}
{\Large \bf
Enhanced ${\cal N}=8$ Supersymmetry}\\
\vskip0.3cm
{\Large \bf of}\\
\vskip0.3cm
{\Large \bf ABJM Theory on $\mathbb{R}^8$ and $\mathbb{R}^8/\mathbb{Z}_2$}\\
\vskip 1.5truecm
{\large
Andreas Gustavsson $^{a,b}$, \,\,\, Soo-Jong Rey $^{a,c}$}\\
\vskip 1.5truecm
{\it $^a$ School of Physics and Astronomy \& Center for Theoretical Physics}
\vskip0.2cm
{\it Seoul National University, Seoul 151-747 {\rm KOREA}}
\vskip0.3cm
{\it $^b$ Center for quantum spacetime (CQUeST), Sogang University, Seoul 121-742 {\rm KOREA}}
\vskip0.3cm
{\it $^c$ Kavli Institute for Theoretical Physics, Santa Barbara
CA 93106 {\rm USA}}
\vskip 0.3cm
{\tt a.r.gustavsson@swipnet.se, \,\,\, sjrey@snu.ac.kr}
\vskip 1.2truecm
{\bf ABSTRACT}
\end{center}
\vskip0.3cm
{The ABJM theory we will study in this paper refers to superconformal Chern-Simons-matter theory with product gauge group $U(N) \times U(M)$ and levels $+k, -k$, respectively. The theory is a candidate for worldvolume dynamics of M2-branes sitting at $\mathbb{C}^4/\mathbb{Z}_k$. By utilizing monopole operators, we prove that ABJM theory exhibits enhanced $\N=8$ supersymmetry and SO(8) R-symmetry at Chern-Simons levels $k=1,2$. We first show that the ABJM Lagrangian can be written in manifestly SO(8) invariant form up to certain extra terms. We then show that upon integrating out Chern-Simons gauge fields these extra terms vanish precisely at levels $k=1,2$. Utilizing monopole operators at these levels, we identify new ${\cal N}=2$ supersymmetry. We demonstrate that they combine with the manifest ${\cal N}=6$ supersymmetry to close on-shell on ${\cal N}=8$ supersymmetry.  We finally show that the ABJM scalar potential is SO(8) invariant.}
\vfill
\vskip4pt
\eject
\pagestyle{plain}

\section{Introduction}
Aharony, Bergman, Jafferis and Maldacena \cite{ABJM} proposed a three-dimensionsl superconformal field theory as a microscopic description for worldvolume dynamics of multiple M2-branes on $SU(4)\times U(1)$ R-symmetric and ${\cal{N}} = 6$ superconformal M2-branes. Hereafter referred as ABJM theory, it is defined by a {\sl gauged} linear sigma model: eight scalar and fermion fields in the bifundamental representation of quiver gauge group $\mathfrak{G} = U(N) \times U(M)$ coupled to Chern-Simons gauge theory. Classically $N$ and $M$ may be arbitrary integers. Quantum mechanically one expects that only theories with $|N-M|\leq |k|$ are unitary \cite{Aharony:2008gk}. For $N=M$ it was proposed in \cite{ABJM} that ABJM theory is holographically dual to the Type IIA string theory on AdS$_4 \times \mathbb{CP}^3$ in the planar limit of both $N$ and $k$ infinite while holding `t Hooft coupling  $\lambda \equiv$ $N/k$ fixed and large. At finite $k$, the holographic dual is described most appropriately by the M theory on AdS$_4 \times \mathbb{S}^7/\mathbb{Z}_k$. The proposal of \cite{ABJM} provides a Type IIA string or M-theory counterpart of the much studied AdS/CFT correspondence \cite{maldacena} between the Type IIB string on AdS$_5 \times \mathbb{S}^5$ and the four-dimensional ${\cal N}=4$ super Yang-Mills theory. Interestingly, there are strong indications that the ABJM theory is integrable, both at weak coupling \cite{weakcouplingintegrable}, \cite{ABJintegrable} and strong coupling \cite{strongcouplingintegrable} regimes.

Built upon this holography, it was further anticipated in \cite{ABJM} that the ABJM theory at Chern-Simons levels $k=1,2$ actually has ${\cal{N}}=8$ supersymmetry and $SO(8)$ R-symmetry which are the symmetries of coincident M2 branes on ${\mb{R}}^{1,2} \times \mathbb{R}^8$ or $\mathbb{R}^{1,2} \times (\mathbb{R}^8/\mathbb{Z}_2)$, respectively. The same should be true for theories with gauge groups $U(N)\times U(M)$ with $N\neq M$. However there are dualities among these theories \cite{Aharony:2008gk}, and the upshot is that there is only one more class of these theories one may consider, namely the ones with gauge groups $U(N)_2\times U(N+1)_{-2}$ at level $k=2$. Any other theory with gauge group $U(N)_k\times U(M)_{-k}$ at levels $k=1,2$ will be either inconsistent at quantum level, or be related by a duality to either one of the theories above. Now $\N=6$ superconformal theories also exist for a few more gauge groups, especially for $Sp(N)\times U(1)$. These theories will get enhanced supersymmetry for other values of the level $k$. To be able to include any $\N=6$ theory in our approach we find it suitable to use the three-algebra formulation of ABJM theory. However when we speak about supersymmetry enhancement for levels $k=1,2$, we will have in mind the special cases of $U(N)\times U(N)$ and $U(N)\times U(N+1)$ gauge groups only. Since our approach will build on three-algebra, everything will also apply to gauge groups $Sp(N)\times U(1)$ with the only modification that the levels $k$ at which we have enhanced supersymmetry are slightly different (we expect only $k=1$ in these cases).

The purpose of this paper is to prove that the ABJM theory with gauge group $U(N)_k\times U(M)_{-k}$, for any possible rank of gauge group, exhibits enhanced ${\cal N}=8$ superconformal symmetry and SO(8) R-symmetry at Chern-Simons level $k=1,2$. Our proof relies on three-algebra structure and monopole operators inherent in this theory. Therefrom, precisely when the Chern-Simons level $k$ takes the value $1$ or $2$, a set of highly nontrivial algebraic identities follows among the matter fields. Utilizing these identities, we prove that the ABJM theory possesses extra ${\cal N}=2$ supersymmetry that combines with the existing ${\cal N}=6$ supersymmetry to the fully enhanced ${\cal N}=8$ supersymmetry and SO(8) R-symmetry.

A feature of the ABJM theory is that the gauge dynamics, governed solely by the Chern-Simons term, is trivial. The Chern-Simons term merely induces braiding statistics to the matter fields. Consequently, operators built solely from the gauge potential such as holonomy and magnetic monopole operators $W_R$ would not carry any dynamics or scaling dimension, though they transform in nontrivial representations $R$ under $\mathfrak{G}$ \cite{kapustin}. Upon coupling matter fields to the Chern-Simons gauge field, gauge invariant operators are constructible not just from matter fields alone but also by attaching the holonomy or magnetic monopole operators $W_R$ to them. Made entirely out of gauge potential, the monopole operators are singlets under internal rigid symmetries such as R-symmetry. As such, monopole operators can produce gauge invariant operators with a rich variety of the R-symmetry representations. Recently, through the study of superconformal index, it was shown that gauge invariant operators containing the monopole operators $W_R$ are indispensable for confirming the AdS/CFT correspondence between the ABJM theory and the M-theory at finite $k$ \cite{seok}.

Another feature of ABJM theory is that high degree of supersymmetry restricts permissible gauge groups, as well as representations of matter contents.
In applications to specific problems, it is useful to formulate the ABJM theory in terms of the Lie algebra $\mathfrak{g}$ of the gauge group $\mathfrak{G}$ and representation $R$ of matter fields. On the other hand, in a formulation that aims at incorporating all possible gauge groups and matter contents compatible with ${\cal N}=6$ supersymmetry, it would be more convenient and unifying to use an algebraic structure that underlies all ABJM theories. It was found in \cite{BLN=6} that the pertinent algebraic structure of the ABJM theory is so-called {\sl hermitian} 3-algebra ${\cal A}_3(\mathbb{C})$.
In this formulation, classification of permissible gauge groups and representations for ${\cal N}=6$ supersymmetry was carried out in \cite{Schnabl:2008wj}. An infinite class of them were found, among which the smallest rank $\mathfrak{G}=$ SO(4)$=$SU(2)$\times$SU(2) is found identical to the Bagger-Lambert-Gustavsson (BLG) theory \cite{BLG}. The BLG theory, however, is known to have {\sl real} 3-algebra ${\cal A}_3(\mathbb{R})$ and ${\cal N}=8$ supersymmetry. This calls for better understanding under what other choices of the ABJM theory parameters would exhibit the maximally enhanced ${\cal N}=8$ supersymmetry and SO(8) symmetry.

Our proof of enhanced symmetries constitutes in showing that, by utilizing the three-algebra ${\cal A}_3(\mathbb{C})$ and the monopole operators $W$, the ABJM theory at Chern-Simons levels $k=1,2$ is expressible as a `trial' BLG theory, where the original real 3-algebra ${\cal A}_3(\mathbb{R})$ is replaced by the hermitian 3-algebra ${\cal A}_3 (\mathbb{C})$. In this way, the ${\cal N}=8$ supersymmetry and the SO(8) R-symmetry become manifest. Here, `trial' refers to the triality of the SO(8) R-symmetry group.

We should point out that, though details differ somewhat, a  variant of the symmetry enhancement at $k=1,2$ works for massive and non-relativistic ABJM theory \cite{NRABJM}  --- the non-relativistic reduction of massive ABJM theory, where only holonomy and monopole operators are known to generate physically nontrivial correlators \cite{nakayamarey}. In fact, this theory illustrates in a clean manner intimate relations among symmetry enhancement between the ABJM and the non-ABJM fields, trivial braiding statistics for $k=1,2$ and bound-states of M-theory momentum modes. However, contrary to a naive extrapolation of consideration of this work, structure of the symmetry algebra indicates that details of the symmetry enhancement should be distinctively different. Related, we also point out that conformal and superconformal invairance do not play central role in our proof of the symmetry enhancement.

This paper is organized as follows. In section 2, we summarize key ideas and provide a roadmap of our proof. In section 3, we illustrate these key ideas and roadmap for abelian gauge group. In section 4, we present details of hermitian 3-algebra ${\cal A}_3(\mathbb{C})$ inherent to the ABJM theory. Also, in section 5, we  present properties of monopole operator. In particular, we pay attention to the general covariance property, which will play a prominent role for foregoing considerations. In section 6, we lay down details of closure among so-called the ABJM fields and the non-ABJM fields -- composites made of the ABJM fields and the rank-2 monopole operators. In section 7, we first identify novel ${\cal N}=2$ supersymmetry that act between the ABJM and the non-ABJM fields. Combining them with the manifest ${\cal N}=6$ supersymmetry yields the maximal ${\cal N}=8$ supersymmetry we are after. In this section, we check explicitly on-shell closure of the ${\cal N}=8$ supersymmetry. In section 8, utilizing the similar reasonings, we show that the ABJM scalar potential is in fact identical to the BLG scalar potential. This demonstrate SO(8) symmetry of the ABJM scalar potential. By ${\cal N}=8$ supersymmetry, the Yukawa interactions also have SO(8) symmetry. In appendix A, we recall SO(8) gamma matrices and several relevant Fierz identities. In appendix B, we also recall SO(1,2) gamma matrices. In appendix C, we summarize branching rule of SO(8) to SU(4)$\times$U(1). In appendix D, we provide Fierz identities of ${\cal N}=6$ superysmmetry, of the new ${\cal N}=2$ supersymmetry and hence of the full ${\cal N}=8$ supersymmetry. In appendix, we explain triality rotated, so-called trial BLG theory.

%%%%%%%%%%%%%%%%%%%%%%%%%%%%%%%%%%%%%%%%%%%%%%%%%%%%%%%%%%%%%%%%%%%%%%%%%%%%%%%%%%%%%%%%%%%%%%%%%%%%%%%%%%%%%
\section{Roadmap and Key Ideas}
In this section, we shall outline key ideas used and a roadmap to our proof.

\subsubsection*{3-algebra}
Since we shall heavily use the 3-algebra formulation throughout, we here summarize its emergence in the BLG and the ABJM theories. As recalled above, underlying algebraic structure of the BLG theory was identified with the {\sl real} 3-algebra ${\cal A}_3(\mathbb{R})$. Its structure constants $f^{bcd}{}_a$ are real-valued and totally antisymmetric in $b,c,d$~\footnote{Note that metric structure of the 3-algebra is not needed for equations of motion and for closing the ${\cal N}=8$ supersymmetry variations, but is imperative for Lagrangian formulation.}. The structure was so restrictive that the only finite-dimensional choice of the gauge group $\mathfrak{G}$ is SU$_L(2)\times$SU$_R(2)=$SO(4). To have more general gauge groups, it became clear one would have to relax the 3-algebra structure. But it seemed impossible to do so while keeping all the global symmetries of the BLG theory intact. A solution to this difficulty was proposed by ABJM \cite{ABJM}, where the SO(8) R-symmetry is given up and only the SU(4)$\times$U(1) part of it is kept manifest. The resulting ABJM theories traded an infinite class of admissible $\mathfrak{G}$ with reduced $\N=6$ supersymmetry and SU(4) R-symmetry.

As recalled above, algebraic structure underlying all admissible ABJM theories is the {\sl hermitian} 3-algebra ${\cal A}_3(\mathbb{C})$ \cite{BLN=6}. Its structure constants $f^{bc}{}_{da}$ are antisymmetric in their two upper and two lower indices, respectively, and hermitian in the sense that
\bea
f^{*bc}{}_{da} &=& f^{da}{}_{bc}.
\eea
In this formulation, we do not need to assume a metric on the 3-algebra since we can use complex conjugation to raise and lower indices \footnote{The hermitian 3-algebra ${\cal A}_3 (\mathbb{C})$ without metric structure can also be found in \cite{palmkvist}.}. Even though we have no metric, we do have a trace-form and we can express the ABJM action using this trace-form. We will refer to the 3-algebra without a metric structure as {\sl{hermitian}} 3-algebra $A_3(\mathbb{C})$ \footnote{The hermitian 3-algebra ${\cal A}_3 (\mathbb{C})$ is a generalization of the real 3-algebra ${\cal A}_3 (\mathbb{R})$. In particular, this also implies that the Nambu 3-bracket is also a realization of the hermitian 3-algebra.}. In this way, all admissible ABJM theories (that includes the BLG theory as one of them) are unified in a single framework of the 3-algebra ${\cal A}_3(\cdot)$.

The classification of \cite{Schnabl:2008wj} may be viewed as a consequence of the hermitian 3-algebra structure and the fundamental identity therein. For ${\cal N}=6$, there is an ABJM theory for every hermitian 3-algebra. A hermitian 3-algebra in turn corresponds to a choice of the gauge group $\mathfrak{G}$ based on a semi-simple Lie group. In this paper, shall we consider ABJM theories that correspond to hermitian 3-algebra, viz. semi-simple Lie group. There can also exists global U(1)$\times$U(1) symmetry, corresponding to conserved baryon numbers, modulo global identifications of center elements. In that case, these U(1)s can be gauged. The resulting theory is the ABJM theory originally proposed \cite{ABJM}.

\subsubsection*{rank-2 monopole operators}
In 3-algebra, we have gauge indices $a,b,... = 1, \cdots, {\rm dim}{\cal A}_3$ associated with 3-algebra generators $T^a$ and their complex conjugates that we denote as $T_a$. The monopole operator that will be useful for us are those with two symmetric gauge indices up or two indices down, $W^{ab} = W^{ba}$ and $W_{ab} = W_{ba}$, respectively. These symmetric rank-2 monopole operators can be used to turn the ABJM scalar field $Z^A_a$ into a field $Z^{Aa} = W^{ab} Z^A_b$ and similarly for the ABJM fermion fields. Here $A$ is an index transforming in the fundamental representation of the global SU(4) R-symmetry of the ABJM theory. With the rank-2 monopole operators at hand, there are two ways to move the 3-algebra indices of the ABJM fields up or down. The first is attaching the rank-2 monopole operator as described above. The second is to take complex conjugate of the ABJM fields. Note that the complex conjugation acts by raising and lowering both gauge and R-symmetry indices, so the scalar field $Z^a_A$ is the complex conjugated field of $Z^A_a$, etc. Summarizing, starting from the matter field $Z^A_a$, we can construct $Z^{Aa}$ or $Z^a_A$ by attaching the monopole operator or by complex conjugation, respectively.

Attaching a monopole operator to a local field renders the composite a non-local operator since the monopole operator depends in general on the Dirac string. If the Dirac-Schwinger-Zwanziger quantization condition is obeyed, the Dirac string is unobservable and the monopole operator becomes a local field configuration. Moreover, the monopole operator is covariantly constant. Below we shall demonstrate this explicitly for the abelian ABJM theory and find that, only for Chern-Simons levels $k=1$ and 2, the composite operators are local field configurations. This fits nicely with the fact that only at levels $k=1,2$ can we expect to have enhanced supersymmetry and R-symmetry. This is our first evidence that monopole operators should play some role in symmetry enhancement of ABJM theory.

\subsubsection*{roadmap}
Denote vector, spinor and cospinor representations of SO(8) as ${\bf 8}_{\rm v}, {\bf 8}_{\rm s}, {\bf 8}_c$, and their basis indices by $I,\alpha,\dot{\alpha} = 1, \cdots, 8$, respectively. In the hermitian BLG theory, matter fields are ${\bf 8}_{\rm v}$ for $X^I_a$ and ${\bf 8}_{\rm s}$ for $\psi_{\alpha a}$. The hermitian BLG theory is then defined by Chern-Simons term and the gauged matter Lagrangian
\bea
\cal{L}_{\rm BLG-v} &=& -\frac{1}{2} D_{\mu} X^{I}_a D^{\mu} X_{I}^a - \frac{1}{12} X^I_b X_I^e X^J_c X_J^f X^K_g X_K^d f^{bc}{}_{da} f^{ga}{}_{ef}\cr
&& + \frac{i}{2} \overline{\psi}^{\alpha a} \gamma^{\mu} D_{\mu} \psi_{\alpha a} + \frac{i}{4}\overline{\psi}^{a\alpha} \Gamma_{I\alpha\dot{\beta}} \Gamma_{J\dot{\beta}\gamma} X_{I b} X_{J c} \psi^{\gamma d} f^{bc}{}_{da} \, .
\eea

We next use the triality of SO(8) group and map the original fields to triality-rotated fields. This way, we can construct two new trial hermitian BLG theories. In all these theories, the Chern-Simons term is universal since it is unaffected by the SO(8) triality. We are interested in the theory obtained by the following triality transformation:
\bea
({\bf 8}_{\rm v}, {\bf 8}_{\rm s}, {\bf 8}_{\rm c}) \rightarrow ({\bf 8}_{\rm s}, {\bf 8}_{\rm c}, {\bf 8}_{\rm v}); \qquad \quad (I,\alpha,\dot{\alpha}) \rightarrow (\alpha,\dot{\alpha},I).
\eea
After the triality rotation, the Lagrangian reads
\bea
\cal{L}_{\rm BLG-s} &=& -\frac{1}{2} D_{\mu} X^{\alpha}_a D^{\mu} X_{\alpha}^a - \frac{1}{12} X^{\alpha}_b X_{\alpha}^e X^{\beta}_c X_{\beta}^f X^{\gamma}_g X_{\gamma}^d f^{bc}{}_{da} f^{ga}{}_{ef}\cr
&&+ \frac{i}{2} \overline{\psi}^{\dot{\alpha}a} \gamma^{\mu} D_{\mu} \psi_{\dot{\alpha}a} + \frac{i}{4}\overline{\psi}^{a \dot{\alpha}} \psi_{\dot{\beta}b} X^{\alpha}_c X_{\beta}^d \Gamma_{I \alpha \dot{\alpha}} \Gamma_I^{\dot{\beta}\beta} f^{bc}{}_{da} \, .  \label{conjugatelagrangian}
\eea
viz. the matter fields are SO(8) spinors and cospinors $\mathbf{8}_{\rm s}, \mathbf{8}_c$ and the supersymmetry is SO(8) vector $\mathbf{8}_{\rm v}$ (see appendix E).
The Lagrangian (\ref{conjugatelagrangian}) is the one related to the ABJM Lagrangian. To show this, we break SO(8) to SO$(6)\times$SO(2)$\simeq$SU$(4)\times$U(1) and decompose the SO(8) spinor and cospinor fields as
\bea
X_{\alpha}^a = \(\begin{array}{c}
Z^A_a\\
Z_{Aa}
\end{array}\), && X^{\alpha}_a = \(\begin{array}{c}
Z_A^a\\
Z^{Aa}
\end{array}\)\cr
\psi_{\dot{\alpha}a} = \( \! \begin{array}{c}
\psi^{A}_a\\
-\psi_{Aa}
\end{array} \! \), && \psi^{\dot{\alpha}a} = \(\! \begin{array}{c}
\psi_{A}^a\\
-\psi^{Aa}
\end{array} \! \) \, .
\eea
We also split the SO(8) gamma matrices into SO(6) and SO(2) gamma matrices as $\Gamma_I = (\Gamma_M,\Gamma_X)$ and denote by $\Sigma_{M,AB}$ and $\Sigma_M^{AB}$ the off-diagonal blocks in $\Gamma_M$. The details are collected in Appendix \ref{gamma}.
The fields $Z^A_a$ and $\psi_{Aa}$ as well as their hermitian conjugates are the ABJM scalar and fermion fields, where upper $A$ is fundamental and lower $A$ is anti-fundamental of SU(4)~\footnote{Equivalently, they are spinor and cospinor of SO(6).}. The fields $Z^{Aa}$ and $\psi_A^a$ are not the ABJM fields --- we refer them as `non-ABJM fields'. Our strategy is to relate the non-ABJM fields to the ABJM fields by means of the monopole operators $W^{ab}, W_{ab}$, since these operators are the unique tensors that can raise or lower indices gauge covariantly.

After the decomposition, the triality-rotated BLG Lagrangian takes the form
\bea
{\cal L}_{\rm matter}
&=&-D_{\mu} Z^A_a D^{\mu} Z_A^a -i\overline{\psi}^{Aa} \gamma^{\mu} D_{\mu} \psi_{Aa} \cr
&& + i \(- \overline{\psi}^{Aa} \psi_{Ab} Z^B_c Z_B^d + 2 \overline\psi^{Ba} \psi_{Ab} Z^A_c Z_A^d \) f^{bc}{}_{da}\cr&&- \(\frac{1}{2} \epsilon^{ABCD} \overline{\psi}_{Bb} Z_{A}^a Z_{D}^d \psi_{Bc} + \frac{1}{2} \epsilon_{ABCD} \overline{\psi}^{Ba} Z^A_b Z^D_c \psi^{Cd} \)f^{bc}{}_{da} \cr
&& - \frac{2}{3}\(f^{ab}{}_{gh} f^{ch}{}_{ef} - \frac{1}{2} f^{ab}{}_{eh} f^{ch}{}_{gf}\) Z^A_a Z_A^e Z^B_b Z_B^f Z^C_c Z_C^g \cr
&& + \cdots \cdots.
\,  \label{abjm}
\eea
The terms shown depend only on the ABJM fields and gives rise precisely to the ABJM Lagrangian. The ellipses denote complicated terms that involve the non-ABJM fields. This brings us to the following interesting question: Under what conditions will the ellipses vanish identically and the trial BLG theory become identical to the ABJM theory? We find that this is so{\sl provided} the following set of algebraic identities hold:
\bea
\(Z^A_c Z_A^d + Z^{Ad} Z_{Ac}\) f^{bc}{}_{da} &=& 0 \nonumber \\
\(Z_{Ab} Z^B_c Z_B^d + Z_A^d Z^B_c Z_{Bb}\) f^{bc}{}_{da} &=& 0 \nonumber \\
\(Z_{Ab} Z_{Bc} Z_C^d - Z_{Cb} Z_{[Ac} Z_{B]}^d\) f^{bc}{}_{da} &=& 0 \nonumber \\
\psi_{Ab} \(Z_{Bc} Z^{Ad} + Z_B^d Z^A_c\) f^{bc}{}_{da} &=& 0 \nonumber \\
\psi_A^b \(Z_B^c Z^A_d + Z_{Bd} Z^{Ac}\) f^{da}{}_{bc} &=& 0 \, . \label{identitiesIntro}
\eea
With these identities, we also find equivalence between the scalar potential in the ABJM theory and the scalar potential in the generalized trial BLG theory, as demonstrated in section \ref{sextic}. Through ${\cal N}=8$ supersymmetry transformations, equivalence between the ABJM and generalized trial BLG Yukawa coupling terms can be checked.

If the ABJM Lagrangian is SO(8) invariant, the identities (\ref{identitiesIntro}) should hold in some sense\footnote{The symmetry enhancement can not be seen in the classical Lagrangian where $k$ is just an overall factor multiplying the whole Lagrangian. But if we integrate out the gauge field then these identities will hold for levels $k=1,2$.} and we can express the ABJM Lagrangian in the manifestly SO(8) invariant form as a generalized trial BLG Lagrangian. We shall show that (\ref{identitiesIntro}) originate from the flatness condition of the gauge field strengths
\bea
\widetilde{F}_{\mu\nu}{}_b{}^a + \widetilde{F}_{\mu\nu}{}^a{}_b &=& 0.\label{fieldstrengthIntro}
\eea
and that the identities (\ref{identitiesIntro}) are all related to (\ref{fieldstrengthIntro}) by $\N=6$ supersymmetry.

Our final step is to discover two extra supersymmetries. We discover them and put together with the ${\cal N}=6$ supersymmetries in SO(8) covariant form.
To show that they give $\N=8$ supersymmetry, it is not enough to just show that the Lagrangian can be written in an SO(8) invariant form. Indeed, we will find that we need a few more identities of a similar type in order to have closure of $\N=8$ supersymmetry transformations modulo the ABJM equations of motion, SO(8) rotation and gauge transformations.

Incidentally, the above algebraic identities may be interpreted as constraining the matter fields\footnote{If we take the viewpoint that the (non-dynamical) gauge field is put on-shell and expressed as a composite field in terms of the matter fields.} $Z^A_a$'s. This may be an indication of the feature of the ABJM theory that the true degrees of freedom scales as $N^{3/2}$, not as $N^2$.
%Notice that these identities are derived from (\ref{fieldstrengthIntro}) using only the ${\cal N}=6$ supersymmetry, so are valid for {\sl any} $k$. As they involve monopole operators, the identities are local identities only for $k=1,2$.

%%%%%%%%%%%%%%%%%%%%%%%%%%%%%%%%%%%%%%%%%%%%%%%%%%%%%%%%%%%%%%%%%%%%%%%%%%%%%%%%%%%%%%%%%%%%%%%%%%%%%%%%%%%
\section{Prelude: abelian ABJM theory}\label{Ab}
\subsection{linear sigma model}
To appreciate the symmetry enhancement clearer, we first study the abelian ABJM theory. Here, of course, the 3-algebra structure ${\cal A}_3(\cdot)$ is not essential. We start with (2+1)-dimensional linear sigma model over the target space $\mathbb{C}^4$. There are four complex scalar fields $Z^A$ and their complex conjugates $(Z^{A})^* = Z_A$. They transform as ${\bf 4}, \overline{\bf 4}$ under SU(4) of the target space. This linear sigma model corresponds to bosonic part of the ABJM theory with gauge group U(1)$\times$U(1) at Chern-Simons level $k=1$, as we will see in the next section. The action reads
\bea
L_{\rm matter} = -\int \rmd^3 x \,\,  \partial_\mu Z^A \partial^\mu Z_A \ .  \label{freeaction}
\eea
The sigma model is invariant under U(4)$=$SU(4)$\times$U(1) transformations:
\bea
\delta Z^A &=& \omega^A{}_B \ Z^B \, ,
\eea
Here,
\bea
(\omega^*)^A{}_B + \omega^B{}_A &=& 0
\eea
are anti-hermitian matrices, generating SU(4) transformations by the traceless parts and U(1) transformation
by the trace part. In total, there are 16 real parameters.

The sigma model (\ref{freeaction}) has more symmetries. It is also invariant under the transformations
\bea
\delta Z^A &=& \omega^{AB} Z_B
\eea
described by 6 complex parameters related by
\bea
\omega^{AB} + \omega^{BA} &=& 0,\cr
\omega^{* \ AB} + \omega_{BA} &=& 0 \, .
\eea
These transformations do not close among themselves. However, when combined with the above SU(4)$\times$U(1) transformations, they are closed and generate the SO(8) symmetry group with $28 = 16 + 6 \cdot 2$ real parameters.

To see the SO(8) symmetry better, we elaborate here somewhat technical but fairly straightforward discussion regarding how part of the SO(8) transformations not contained in SU(4)$\times$U(1) acts on ${\bf 8}_{\rm v}$ and ${\bf 8}_{\rm s}$ representations of SO(8). The results obtained here will be useful later. Acting on a ${\bf 8}_{\rm v}$ representation $V_I$ ($I=1,...,8$), an infinitesimal SO(8) transformation is given by
\bea
\delta V_I &=& \omega_{IJ} V_J
\eea
where $\omega_{IJ}$ is anti-hermitian and has real components (in other words, it is antisymmetric). We decompose ${\bf 8}_{\rm v}$ into a six-dimensional part $V^M$ ($M=1,...,6$) and a two-dimensional part $V = v^7+i v^8$. The metric being Kronecker deltas, we do not distinguish upper or lower SO(8) or SO(6) indices. The SO(2) parameter is $\omega^{78}$ and the SO(6) parameters are $\omega^{MN}$. We are mainly interested in the SO(8) rotations that mix SO(6) with SO(2). These rotations are parametrized by $\omega^M := \omega^{M7}+i\omega^{M8}$ and act on the SO(8) vector as
\bea
\delta V^M &=& \frac{1}{2} \(\omega^M \ V^* + \omega^{*M} \ V \)\cr
\delta V \ &=& \omega^M \ V^M\cr
\delta V^* &=& \omega^{*M} \ V^M \, .
\eea
An SO(8) Dirac spinor decomposes into Weyl $X_{\alpha}$ and anti-Weyl spinor $\psi_{\dot{\alpha}}$. These in turn decompose into Weyl spinors of SO(6). We define these Weyl components as
\bea
X_{\alpha} &=& \(\begin{array}{c}
Z^A\\
Z_A
\end{array}\)
\eea
\bea
\psi_{\dot{\alpha}} &=& \(\begin{array}{c}
\psi^A\\
-\psi_A
\end{array}\).
\eea
On the SO(8) R-symmetry Dirac spinor\footnote{It is important that this is R-symmetry spinor as opposed to spacetime spinor. In particular, $Z$ is commuting bosonic field.}
\bea
\Xi &=& \(\begin{array}{c}
X_{\alpha}\\
\psi_{\dot{\alpha}} \ ,
\end{array}\)
\eea
an infinitesimal SO(8) transformation acts as
\bea
\delta \Xi &=& -\frac{1}{2} \omega^{MX} \Gamma^{MX} \Xi.
\eea
Here, the normalization is fixed by how the vector index of gamma matrices transforms (as a direct consequence of the Clifford algebra),
\bea
[\Gamma_{IJ},\Gamma_K] &=& -4 \delta_{K[I}\Gamma_{J]}.
\eea
One can view this as the invariance condition of the gamma matrices where all its indices are transformed. Explicitly, we find the variations as
\bea
\delta Z^A &=& \frac{i}{2} \omega^M \Sigma^{M,AB} Z_{B}\cr
\delta Z_A &=& \frac{i}{2} \omega^{*M} \Sigma^M_{AB} Z^{B},
\eea
\bea
\delta \psi^{A} &=& \frac{i}{2} \omega^{*M} \Sigma^{M,AB} \psi_B\cr
\delta \psi_{A} &=& \frac{i}{2} \omega^M \Sigma^M_{AB} \psi^B.
\eea

%%%%%%%%%%%%%%%%%%%%%%%%%%%%%%%%%%%%%%%%%%%%%%%%%%%%%%%%%%%%%%%%%%%%%%%%%%%%%%%%%%%%%%%%%%%%%%%%%%%%%%%
\subsection{gauging U(1) symmetry}\label{Ga}
\subsubsection*{Chern-Simons gauging:}
We now gauge the U(1) symmetry by introducing a {\sl flat} one-form gauge field $b$. We then define the covariant derivative
\bea
D Z^A := \rmd Z^A+ib Z^A
\eea
and consider the gauged linear sigma model
\bea
-\int \rmd^3 x \,\,\, \(D_{\mu} Z^A D^{\mu} Z_A + \frac{k}{2\pi} b\wedge  \rmd \ a\) .
\eea
Here, $a$ is a Lagrange multiplier one-form gauge field that constrains $b$ to be flat, $\rmd b = 0$. This model equals to the bosonic part of the abelian ABJM action at integer-valued Chern-Simons level $k$.

We can integrate out $a$, setting \cite{ABJM, Schwarz}
\bea
k \ b = \rmd \ \sigma \ .
\eea
This gives back the linear sigma model modulo the orbifold identification
\bea
Z^A \simeq e^{\frac{2\pi i}{k}} Z^A.
\eea
In general, this identification breaks SO(8) down to SU(4)$\times$U(1). At $k=1,2$, however, the
SO(8) symmetry is retained. If the $\mathbb{Z}_k$ orbifolding is SO(8) invariant, it should commute with
the transformation
\bea
Z^A \rightarrow Z^A + \omega^{AB} \ Z_B \ .
\eea
This implies that
\bea
Z^A \rightarrow Z^A + \omega^{AB} e^{-\frac{4\pi i}{k}} Z_B
\eea
should also be a symmetry. This singles out the Chern-Simons coefficient $k$ to $1, 2$.

\subsubsection*{monopole operators:}
Notice that SO(8) symmetry cannot act in this simple way were the gauge field not integrated out.
The transformation
\bea
Z^A \rightarrow Z^A + \omega^{AB} Z_B
\eea
would not be gauge covariant since $Z^A$ and $Z_A$ are oppositely charged with respect to the gauge field $b$. The remedy for this is to redefine the scalar fields by attaching monopole operators to these fields in such a way that all equations transform covariantly under the U(1) gauge transformations. At level $k$, the monopole operator that we have at our disposal is of the form
\bea
T_k &=& e^{i\sigma}.
\eea
From the Chern-Simons term, we also see that this operator carries $k$ unit of electric charge.
Thus, the gauge transformations act as
\bea
T_k & \rightarrow & e^{ik\alpha} T_k\cr
Z^A & \rightarrow & e^{i\alpha} Z^A\cr
Z_A & \rightarrow & e^{-i\alpha} Z_A \ .
\eea
At level $k=1$, we can make the field redefinitions
\bea
Z^A & \rightarrow & Z^A\cr
Z_A & \rightarrow & T_1 T_1 Z_A .
\eea
At level $k=2$, we can also make the field redefinitions
\bea
Z^A & \rightarrow & Z^A\cr
Z_A & \rightarrow & T_2 Z_A.
\eea
On these redefined fields, the SO(8) transformation acts in a gauge covariant way. Important observation is that, for $k > 2$, no such local field redefinition is possible. Therefore, this is another way to see that we can have enhanced SO(8) symmetry only for $k=1, 2$.

The Chern-Simons coefficient $k=1,2$ is also special for a seemingly different reason. Consider two external probes charged electrically under the gauge fields $a$ and the $b$, respectively. Upon encircling one of the probes around the other once, we pick up the Aharonov-Bohm phase $\exp(2 \pi i / k)$ as braiding statistics. For $k=1$, the phase is trivial and braiding statistics is bosonic. For $k=2$, the phase is $\pi$ and braiding statistics is fermionic. For $k > 2$, the braiding statistics is anyonic. By the same argument,
we see that the composite we formed above would retain the field statistics unchanged for $k=1, 2$ but
not so for $k >2$.

\subsubsection*{local versus nonlocal:}
The reason we have these monopole operators at our disposal comes from the Chern-Simons action. Consider the monopole operator
\bea
\exp i\sigma(x)  := \exp  \( i \int_\infty^x \rmd \sigma (x) \).
\eea
Naively, one could think that operators of the form $\exp (i\sigma(x)/\ell)$ is also feasible, where $\ell$ is an arbitrary integer. However, this is not so because $\sigma$ is a compact pseudo-scalar defined over the period $2\pi$. This means that that $\oint \rmd\sigma/\ell \simeq (2\pi/\ell) {\mb{Z}}$ when we integrate over a closed contour. Therefore, $\exp \int \rmd\sigma / \ell$ will be path-dependent, and hence non-local {\sl unless} $\ell =1$.

Not only being local, the monopole operator or products of it is also covariantly constant. Recalling that the monopole operator $T_k$ carries an electric charge of $k$ unit, the covariant derivative acting on it is defined by
\bea
D_{\mu} T_k = \(\partial_\mu - i k b_\mu \) \ T_k = \(i\partial_{\mu} \sigma - i k b_{\mu}\) \ e^{i\sigma}\, .
\eea
We see that this indeed vanishes $D_\mu T_k = 0$ by the defining relation of the dual scalar field, $ k b = \rmd \sigma$. This shows that $T_k$ is covariantly constant. Notice that this property holds for {\sl any} $k$.
% This leads to $D_\mu W_{a_1 \cdots a_k} = 0$ in nonabelian case.

Using these properties, we can put $Z^A$ and $Z_A$ fields on equal footing by attaching appropriate monopole
operators to them. So, $Z^A$ carries an electric charge of one unit, while $(T_k)^n Z_A$ carries an electric charge of $nk-1$. From the above analysis, we see that these two (composite) fields are {\sl local} operators and, as discussed above, can carry equal electric charge when $k=1$ and $n=2$ or $k=2$ and $n=1$, but none for $k > 2$.

%%%%%%%%%%%%%%%%%%%%%%%%%%%%%%%%%%%%%%%%%%%%%%%%%%%%%%%%%%%%%%%%%%%%%%%%%%%%%%%%%%%%%%%%%%%%%%%%%%%%%%%%%
\section{The ABJM theory}
%%%%%%%%%%%%%%%%%%%%%%%%%%%%%%%%%%%%%%%%%%%%%%%%%%%%%%%%%%%%%%%%%%%%%%%%%%%%%%%%%%%%%%%%%%%%%%%%%%%%%%%%
\subsection{hermitian 3-algebra}
The ABJM theory is isomorphic to Hermitian 3-algebras up to possible $U(1)$ factors in the gauge group.
As said, instead of studying the ABJM theory for each possible gauge group separately, it is convenient to utilize the 3-algebra formulation that puts all the possible gauge groups on equal footing. The only property of the gauge groups we need is then the corresponding fundamental identity of the 3-algebra.

\subsubsection*{so(4):}
The simplest example of a 3-algebra is that of gauge group $\mathfrak{G}=$SU$_L(2) \times$SU$_R(2)=$SO(4). This corresponds to a real (which of course also is hermitian) 3-algebra. To see this, we note the following gamma matrix identity among the SO(4) gamma matrices $\gamma_a$ and the chirality matrix $\gamma$:
\bea
\gamma_a \gamma_c \gamma_b - \gamma_b \gamma_c \gamma_a &=& 2 \epsilon_{abcd} \gamma \ \gamma_d.
\eea
In the Weyl representation, the 3-algebra generators $T^a$ sit in the gamma matrices as
\bea
\gamma_a &=& \(\begin{array}{cc}
0 & (T^a)^{i'}_i\\
(T_a)^j_{j'} & 0
\end{array}\)
\eea
Here upper (lower) indices $i$ and $i'$ are (anti)fundamental of SU$_L(2)$ and SU$_R(2)$, respectively. The gamma matrix identity above amounts to the 3-algebra
\bea
T^a T_c T^b - T^b T_c T^a &=& f^{ab}{}_{cd} T^d
\eea
with real structure constants $f^{ab}{}_{cd} = 2\epsilon_{abcd}$. Note that SO(4) also happen to have the metric $\delta_{ab}$ that we can use to raise and lower indices. It is related to the epsilon tensors of SU$_L(2) \times$SU$_R(2)$ as
\bea
\delta_{ab} (T^a)^i_{i'} (T^b)^j_{j'} &=& 2 \epsilon^{ij} \epsilon_{i'j'}.
\eea
We also have
\bea
(T^a)^i_{i'} (T_a)^{j'}{}_j &=& 2 \delta^i_j \delta^{j'}_{i'}.
\eea
For generic ABJM gauge groups there is no such invariant tensor that we can use to raise and lower indices. What we can use instead are monopole operators.

\subsubsection*{generalizations:}
We now generalize the SO(4) 3-algebra by keeping some of the structure of it but dropping the constraints of having real structure constants and a metric. We denote the complex 3-algebra generators by $T^a$. We define complex conjugation as
\bea
T^{*a} &=& T_a.
\eea
The 3-bracket maps three elements into a new element
\bea
[T^a,T^b;T^c] &=& f^{ab}{}_{cd} T^d.
\eea
Here, $f^{ab}{}_{cd}$ are complex-valued structure constants of the 3-algebra. The 3-bracket has the properties
\bea
[T^a,T^b;T^c] &=& -[T^b,T^a;T^c]\cr
[\lambda T^a,T^b;T^c] &=& \lambda [T^a,T^b;T^c]\cr
[T^a,T^b;\lambda T^c] &=& \lambda^* [T^a,T^b;T^c].
\eea
The 3-bracket obeys the so-called fundamental identity. The fundamental identity is best understood as a property of the derivation
\bea
\delta := [\cdot,T^b;T^a]\omega^a{}_b, \label{derivation}
\eea
Here, $\omega^a{}_b$ is an arbitrary anti-hermitian matrix:
\bea
\omega^{*a}{}_b &=& -\omega^b{}_a \, .
\eea
The derivation property is
\bea
\delta [T^e,T^d;T^c] &=& [\delta T^e,T^d;T^c] + [T^e,\delta T^d;T^c] + [T^e,T^d;\delta T^c].
\eea
Using (\ref{derivation}), this amounts to the fundamental identity:
\bea
&& [[T^e,T^d;T^c],T^b;T^a] \nonumber \\
&=& [[T^e,T^b;T^a],T^d,T^c] + [T^e,[T^d,T^b;T^a];T^c] - [T^e,T^d;[T^c,T^a;T^b]].\label{fi}
\eea
In terms of the structure constants, the identity reads
\bea
f^{ed}{}_{cf} f^{fb}{}_{ag} &=& f^{eb}{}_{af} f^{fd}{}_{cg} + f^{db}{}_{af} f^{ef}{}_{cg} - f^{*ca}{}_{bf} f^{ed}{}_{fg}.
\eea

\subsubsection*{inner product:}
We also introduce inner product $\<\cdot,\cdot\>$ such that
\bea
\<T^a,T^b\> &=& \delta^a_b\cr
\<T^a,T^b\> &=& \<T^b,T^a\>^*\cr
\<T^a,T^b\> &=& \< T^{}_b,T^{}_a \> \label{innerproduct}
\eea
By expanding a field $X$ in the 3-algebra basis $X = X_a T^a$, the last property can also be phrased as
\bea
\<X,Y\> &=& \<Y^{*},X^{*}\> \qquad \mbox{for} \qquad X = X_a T^a, \quad Y= Y_a T^a \ .
\eea
This may be taken as defining equation of the hermitian conjugate. Moreover, the inner product has the invariance property
\bea
\<\delta T^a,T^b\> + \<T^a,\delta T^b\> &=& 0
\eea
Using (\ref{derivation}), we get
\bea
f^{*ab}{}_{cd} &=& f^{cd}{}_{ab} \ .\label{hermitian}
\eea
One can also check that this condition can be written as
\bea
\<X,[Y,Z;U]\> &=& \<[X,U;Z],Y\> \ .\label{trinv}
\eea
We note that (\ref{fi}), (\ref{trinv}) generalize the corresponding equations for totally antisymmetric 3-brackets introduced originally for the BLG theory. To get the corresponding fundamental identity and inner product invariance condition for totally antisymmetric 3-bracket, we just need to replace $[\cdot,\cdot;\cdot]$ by totally antisymmetric 3-bracket $[\cdot,\cdot,\cdot]$.

\subsection{matrix realization of hermitian 3-algebra}
\subsubsection*{matrix realization:}
A matrix realization of the 3-algebra ${\cal A}_3(\cdot)$ is provided by
\bea
[X,Y;Z] &:=& X Z^{\dag} Y - Y Z^{\dag} X\cr
\<X,Y\> &:=& \tr(X Y^{\dag}) \label{matrixrealization}.
\eea
The matrix-valued fields $X, Y, Z$ are expanded as $X = X_a T^a$ etc., where $T^a$ is a basis of $(M\times N)$ matrices and $T_a$ are their hermitian conjugates. The 3-bracket is then a map from $M\times N$ matrices to itself -- the first requirement of an algebra. Moreover, the bracket satisfies the fundamental identity (\ref{fi}). Hence, it is a realization of the 3-algebra ${\cal A}_3(\cdot)$, called the Lie 3-algebra ${\cal A}_3 (\mathfrak{g})$.

An explicit solution to the fundamental identity can also be realized in terms of the generators $t^\alpha$ of the associated semi-simple Lie algebra $\mathfrak{g}$ as \cite{BLN=6}
\bea
f^{ab}{}_{cd} &=& (t^{\alpha})^a{}_d (t_{\alpha})^b{}_c
\eea
where $(t^{\alpha})^a{}_b$ are the generators in the bi-fundamental representation.
The index $\alpha$ is lowered by the inverse of Killing form $\kappa^{\alpha\beta}$ of the Lie algebra $\mathfrak{g}$. This realization does not in general satisfy antisymmetry with respect to $a,b$ or $c,d$ indices. Imposing this property restricts possible choices of the Lie algebras $\mathfrak{g}$ and hence the Lie group $\mathfrak{G}$. With the Lie group $\mathfrak{G} = G_L \otimes G_R$, $a,b,c,d$ ranges over $1, \cdots, \mbox{rank}(G_L) \mbox{rank}(G_R)$ and $\alpha$ ranges over $1, \cdots, \mbox{dim}(G_L) + \mbox{dim}(G_R)$. Recall that the index $\alpha$ is lowered by the inverse of the Killing form $\kappa^{\alpha \beta}$ on the semi-simple Lie algebra.

\subsubsection*{similarity transformations:}
We can consider two types of similarity transformations of the Lie algebra generators associated with the 3-algebra. The first type is
\bea
(t^{\alpha})^a{}_b &\rightarrow & {U^a}_c {(t^{\alpha})^c}_d {U^{\dag d}}_b \cr
&\equiv & {\cal U}^{\alpha}{}_{\beta} (t^{\beta})^a{}_b
\eea
where ${U^a}_b {U^{\dag b}}_c = \delta^a_c$ and ${U^{\dag b}}_c {U^c}_a = \delta^c_a$. The second type is
\bea
(t^{\alpha})^a{}_b &\rightarrow & U_{bc} (t^{\alpha})^c{}_d U^{da}
\eea
where $U^{ab}U_{bc} = \delta^a_c$ and $U_{ab} = U_{ba}$.
Both types of transformations leave the Killing form $\kappa^{\alpha\beta}$ invariant, and hence the 3-algebra structure constants are invariant. Explicitly,
\bea
f^{ab}{}_{cd} &=& f^{ef}{}_{gh} \ {U^a}_e {U^b}_f {U^{\dag g}}_c {U^{\dag h}}_d \label{prop1}
\eea
and
\bea
f^{ab}{}_{cd} &=& f^{ef}{}_{gh} \ U^{ga} U^{hb} U_{ce} U_{df} \ , \label{prop2}
\eea
respectively.
Notice that the first type of transformations form a closed group, while the second is not. However,
the total sum of the two types again forms a closed transformation group, which we denote as $\widehat{G}$.

The first type of similarity transformation means that the 3-algebra is invariant under the unitary transformation
\bea
T^a &\rightarrow & T^b {U^a}_b
\eea
The infinitesimal version of this invariance condition leads to the fundamental identity. Namely if we write
\bea
{U^a}_b = \delta_b^a + \Omega^a{}_b
\eea
we find that
\bea
\delta f^{bc}{}_{da} &=& 0   \label{fundidentityinf}
\eea
where we define
\bea
\delta f^{bc}{}_{da} &=& \Omega^b{}_e f^{ec}{}_{da} + \Omega^c{}_e f^{be}{}_{da} - \Omega^e{}_d f^{bc}{}_{ea} - \Omega^e{}_a f^{bc}{}_{de}.
\eea
To make the connection with the fundamental identity, we just write out $\Omega^b{}_a = \omega^d{}_c f^{bc}{}_{da}$.

The second type of similarity transformation is the transformation we shall use repeatedly in later sections.

%%%%%%%%%%%%%%%%%%%%%%%%%%%%%%%%%%%%%%%%%%%
\subsection{ABJM theory in hermitian 3-algebra}
We now describe the ABJM theory in 3-algebra formulation and arrive at (\ref{abjm}).

\subsubsection*{lagrangian:}
In 3-algebra formulation, the covariant derivative is given by
\bea
i D_{\mu} Z_a &:=& i \partial_{\mu} Z_a + Z_b \t A_{\mu}{}^b{}_a; \qquad D_\mu \psi_a := \partial_\mu \psi_a + \psi_b \t A_{\mu}^b{}_a,
\eea
where
\bea
\t A_{\mu}{}^b{}_a \equiv A_{\mu}{}^d{}_c f^{bc}{}_{da}.
\eea
Our gauge fields are anti-Hermitian:
\bea
\t A^*_\mu{{}^b}_a = - \t A_\mu{{}^a}_b \qquad \mbox{equivalently} \qquad  A^*_\mu{}^b{}_a = - {{ A_\mu}^a}_b.
\eea
To translate the action to the more familiar Lie algebra formulation, we use some properties of the 3-algebra of the previous subsection. We just use the matrix realization (\ref{matrixrealization}). We also define gauge fields of the two Lie groups $G_L, G_R$ associated with the 3-algebra by
\bea
A^L_{\mu} &=& A_{\mu}{}^d{}_c T^c T_d\cr
A^R_{\mu} &=& A_{\mu}{}^d{}_c T_d T^c \, .
\eea
With these steps, we find the followings. First, the Chern-Simons term in the 3-algebra formulation turns into two Chern-Simons terms in Lie algebra formulation:
\bea
{k \over 2 \pi} \epsilon^{\mu \nu \lambda} {\rm Tr} (A^L_\mu \partial_\nu A^L_\lambda + {2 i \over 3} A^L_\mu A^L_\nu A^L_\lambda) -
{k \over 2 \pi} \epsilon^{\mu \nu \lambda} {\rm Tr} (A^R_\mu \partial_\nu A^R_\lambda + {2 i \over 3} A^R_\mu A^R_\nu A^R_\lambda) \ .
\eea
Second, the gauge covariant derivatives acting on matter fields are given by
\bea
i D_{\mu} Z^A &=& i \partial_{\mu} Z^A - A^L_{\mu} Z^A + Z^A A^R_{\mu}
\eea
and similarly for fermions. Third, the Yukawa-like terms are given by
\bea
\overline{\psi}^{Aa} \psi_{Ab} Z^B_c Z_B^d &=& {\rm Tr} (\overline{\psi}^A \psi_A Z_B Z^B) - {\rm Tr} (\overline{\psi}^A Z^B Z_B \psi_A)
\eea
etc. The same works for the scalar potential terms. This shows that the ABJM action (\ref{abjm}) in
3-algebra formulation is identical to the ABJM action in Lie algebra formulation, as demonstrated first in \cite{BLN=6}.

\subsubsection*{on-shell ${\cal N}=6$ supersymmetry:}
For later use, we here enlist ${\cal N}=6$ supersymmetry transformations of the ABJM theory in the 3-algebra formulation. They are
\bea
\delta Z^A_a \ \ &=& -i\overline{\epsilon}^{AB} \psi_{Ba}\cr
\delta \psi_{Aa} \ &=& \gamma^{\mu} \epsilon_{AB} D_{\mu} Z^B_a - \(\epsilon_{AB} Z^B_b Z^C_c Z_C^d + \epsilon_{BC} Z^B_b Z^C_c Z^d_A\) f^{bc}{}_{da}\cr
\delta \t A_{\mu}{}^b{}_a &=& \Big(i \overline{\epsilon}_{AB}\gamma_{\mu}Z^A_c\psi^{Bd} -i\overline{\epsilon}^{AB} \gamma_{\mu} \psi_{Ac} Z_B^d \Big)f^{bc}{}_{da}.\label{ABJM}
\eea
The closure relations read
\bea
[\delta_{\eta},\delta_{\epsilon}] Z^A_a \ \ &=& -2i \overline{\epsilon}^M \gamma^{\mu} \eta^M D_{\mu} Z^A_a + \t \Lambda^b{}_a Z^A_b,\cr
[\delta_{\eta},\delta_{\epsilon}] \psi_{Aa} \ &=& -2i \overline{\epsilon}^M \gamma^{\mu} \eta^M D_{\mu} \psi_{Aa} + \t \Lambda^b{}_a \psi_{Ab}\cr
&& + i\overline{\epsilon}^M \gamma^{\lambda} \eta^M \gamma_{\lambda} E_{Aa} + i \overline{\epsilon}^{M}(\Sigma^{MN})_A{}^B \eta^{N} E_{Ba},\cr
[\delta_{\eta},\delta_{\epsilon}] \t A_{\mu}{}^b{}_a &=& -2i \overline{\epsilon}^M\gamma^{\mu} \eta^M \t F_{\nu\mu}{}^b{}_a - D_{\mu} \t \Lambda{}^b{}_a
\eea
with the gauge parameter
\bea
\t \Lambda^b{}_a &=& 2i\overline{\epsilon}^M (\Sigma^{MN})_A{}^B \eta^N Z^A_c Z_B^d f^{bc}{}_{da}.
\eea
The equations of motion needed to close the supersymmetry on-shell are $E_{Aa} = 0$ with
\bea
E_{Aa} &=& \gamma^{\mu} D_{\mu} \psi_{Aa} + \(\psi_{Ab} Z^C_c Z_C^d - 2 \psi_{Bb} Z^B_c Z_A^d + \epsilon_{ABCD} Z^B_b Z^C_c \psi^{Dd}\) f^{bc}{}_{da}\label{fermion}
\eea
for the fermions and
\bea
\t F_{\mu\nu}{}^b{}_a &=& -\epsilon_{\mu\nu\lambda}\(Z^A_c D^{\lambda} Z_A^d - D^{\lambda} Z^A_c Z_A^d-i\overline{\psi}^{Ad}\gamma^{\lambda}\psi_{Ac}\)f^{bc}{}_{da}\label{gaugefield}
\eea
for the gauge field.

%%%%%%%%%%%%%%%%%%%%%%%%%%%%%%%%%%%%%%%%%%%%%%%%%%%%%%%%%%%%%%%%%%%%%%%%%
\section{Monopole Operator and Gauge Covariance}\label{Wilson}
In this section, we shall introduce a monopole operator of requisite property which will play a central role in the foregoing discussions. Consider for definiteness the gauge group $SU$(M) $\times$ SU($N$). We start with infinitesimal gauge transformations
\bea
\delta \t A_{\mu}{}^b{}_a &=& - D_{\mu} \t \Lambda^b{}_a\cr
\delta Z^A_a \ &=& \ Z_b \t \Lambda^b{}_a
\eea
on gauge field and matter fields,  respectively, where
\bea
\t \Lambda^b{}_a &=& \Lambda^c{}_d f^{bc}{}_{da}.
\eea
and $\Lambda^c{}_d$ is any antihermitian matrix.

%In the Lie algebra formulation, the gauge transformation with gauge group element $(g^L,g^R)$ acts on a bifundamental matter field as
%\bea Z \quad \rightarrow \quad g^L Z {g^R}^{\dag} .  \eea

%We introduce monopole operators transforming in the fundamental representation of $G_L, G_R$: \bea W^L &\rightarrow W^L {g^L}^{\dag}\cr W^R &\rightarrow W^R {g^R}^{\dag}. \eea Then the combination
%\bea W^L Z {W^R}^{\dag} &=& Z_a W^L T^a {W^R}^{\dag} \eea is a gauge singlet. Of course this operation does not bring the matter field outside the 3-algebra, so this must again be some linear combination of 3-algebra generators. We define
%\bea W_{a'}{}^a T^{a'} &\equiv & W^L T^a {W^R}^{\dag}. \eea We also define \bea Z_{a'} = W_{a'}{}^a Z_a \eea
%and this will be gauge singlet. We use the prime on indices (like $a'$) to indicate gauge singlet indices at infinity. At infinity gauge parameter drops to zero and indices at infinity do not transform under local gauge transformations.

The scalar fields in the Lie algebra and the 3-algebra basis are related by
\bea
Z^i_{\alpha} = Z_a (T^a)^i_{\alpha}
\eea
and similarly for the fermion fields.
Here $i, \alpha$ are indices of ${\bf M}, \overline{\bf N}$, respectively. Complex conjugate field is
\bea
(Z^*)^i_{\alpha} = Z_i^{\alpha} = Z^a (T_a)_i^{\alpha} \ .
\eea
Gauge transformation with gauge group element $(g^L,g^R)$ acts on the bi-fundamental matter field as
\bea
Z^i_{\alpha} &\rightarrow & (g^L)^i{}_j Z^j_{\beta} ({g^R}^{\dag})^{\beta}{}_{\alpha} .
\eea

\subsection{nonabelian monopole operators}
We now introduce monopole operators \cite{kapustin}. For nonabelian gauge groups, following Goddard-Nuyt-Olive, we may define the monopole operator by embedding the abelian magnetic monopoles to the nonabelian gauge group $\mathfrak{g}$. For each Cartan U(1) subgroup, we specify the abelian magnetic monopole configuration:
\bea
F = {1 \over 4 \pi r^2} \mbox{diag} ( \ m_1, \ m_2, \cdots, \ m_d \ ) \
\eea
up to $\mathfrak{g}$ conjugation. Classically, monopole operators have vanishing engineering dimension and R charge. Quantum mechanically, they may be afflicted by anomalous contributions. We shall assume that this is not the case, as was done recently in \cite{seok}. See also \cite{Klebanov} for a further recent work.

Recall that `t Hooft loop operator is defined by the operation of singular gauge transformations around a (closed) contour $C$. The operation creates a magnetic flux at and along the contour $C$. In a theory with Chern-Simons coupling such as BLG or ABJM, the gauge transformation gives rise to Wilson loop operator along $C$. This implies that the Chern-Simons coupling equates `t Hooft loop operator to the Wilson loop operator. Now, introduce a magnetic monopole. The monopole can open up the contour $C$ of the `t Hooft operator into the monopole operator, connected by semi-infinite `t Hooft operator. The above equivalence between the `t Hooft and the Wilson loop operators is then extendible to the equivalence between the flux-creating monopole operator defined on a semi-infinite contour $C_o$ and the charge-creating holonomy operator defined on the same semi-infinite contour $C_o$. If the Chern-Simons level is $k$ and the monopole charge is $( \ m_1, m_2, \cdots, \ m_d \ )$, the holonomy operator is in the representation of $(k m_1, k m_2, \cdots, k m_d)$ Young tableaux.
%Notice that magnetic monopole alone is incompatible with Chern-Simons term since the latter induces topological mass to the gauge bosons.

The monopole operator that transforms in the fundamental representations of $G_L=$U$(M)$ and $G_R=$U$(N)$ are denoted as $W^L$ and $W^R$, respectively:
\bea
(W^L)_i &\rightarrow (W^L)_j ({g^L}^{\dag})^j{}_i\cr (W^R)_{\alpha} &\rightarrow (W^R)_{\beta} ({g^R}^{\dag})^{\beta}{}_{\alpha}.
\eea
Utilizing them, it is possible to obtain composite fields transforming differently. For example, one can form a gauge singlet composite of the bi-fundamental field $Z$ and monopole operators:
\bea (W^L)_i Z^i_{\alpha} ({W^R}^{\dag})^{\alpha} &=& Z_a (W^L)_i (T^a)^i_{\alpha} ({W^R}^{\dag})^{\alpha}
\ . \eea
Obviously, such an operation does not bring the matter field outside the 3-algebra ${\cal A}_3$, so the composite must again be some linear combination of 3-algebra generators. As such, we define the monopole operator of defining representation in 3-algebra formulation as
\bea W^a &\equiv & (W^L)_i (T^a)^i_{\alpha} ({W^R}^{\dag})^{\alpha} .\eea
Therefore,
\bea Z = W^a Z_a \eea
will be the above gauge singlet composite operator. Associated with $W^a$, there is also the monopole operator
$W_a = W^{*a}$ transforming in the complex conjugate representation.

We can also form composites of other representations than the bi-fundamental, but again the resulting composite operator must be some linear combination of 3-algebra generators. In fact, in order to extend $\N=6$ supersymmetry to $\N=8$ supersymmetry, we may need the monopole operators of higher representations \cite{Klebanov:2008vq}. The most general monopole operator in the Lie algebra and in the 3-algebra basis are related each other as
\bea
W^{a_1...a_k} &=& W^{\alpha_1...\alpha_k}_{i_1...i_k} (T^{a_1})^{i_1}_{\alpha_1}...(T^{a_k})^{i_k}_{\alpha_k}.
\eea
It turns out sufficient to consider symmetric rank-2 representations, $W^{ab}$ and $W_{ab}$. We note that these monopole operators can act to lower and raise gauge indices in a covariant way. For example, by attaching these monopole operators, we have
\bea
Z^{Aa} = W^{ab} Z^A_b, \qquad Z_{Aa} = W_{ab} Z_A^b \ .
\eea
Beware these operations are different from complex conjugation $Z^{A*}_a = Z_A^a$ etc. In particular, the SU(4) representation is not affected by attaching the monopole operators.

Under gauge transformations, the rank-2 monopole operators transform as
\bea
\delta W^{ab} &=& - W^{cb} \t \Lambda_c{}^a -  W^{ac} \t \Lambda_c{}^b\cr
\delta W_{ab} &=& \t \Lambda_a{}^c W_{cb} +   \t \Lambda_b{}^c W_{ac}\label{Wgauge}
\eea
Moreover, they have the properties
\bea
W_{ac}W^{cb} &=& \delta_a^b\cr
W_{ab} &=& W_{ba}\cr
W^{ab} &=& W^{ba}
\eea
In the Lie algebra formulation, the relevant monopole operator is the one in bi-fundamental representations
\bea
W^\alpha_i &=& (W^{R\dag})^\alpha (W^L)_i\cr
W^i_\alpha &=& (W^{L\dag})^i (W^R)_\alpha.
\eea
They are related to the rank-2 monopole operators $W^{ab}, W_{ab}$ by
\bea
W^{ab} T_b &=& W T^a W\cr
W_{ab} T^b &=& W^{\dag} T_a W^{\dag}.\label{eqs}
\eea

\subsection{general covariance}
So far, we focused primarily on the representation contents of the monopole operators.
In general, the monopole operators of a given representation are nonlocal. For the symmetric rank-2
representations, by the Dirac quantization condition, the monopole operator turns out a local operator only
if the Chern-Simons level takes values $k=1$ or $2$. This locality condition leads to an important condition to the gauge field strength, which plays an essential role in foregoing considerations concerning supersymmetry enhancement. Much like the abelian case, invisibility of Dirac string implies that the monopole operator is covariantly constant:
\bea
D_\mu W_{cb} \equiv \partial_{\mu} W_{cb} + \t A_{\mu}^d{}_c W_{db} + \t A_{\mu}{}^d{}_b W_{cd} &=& 0.
\label{covconst}
\eea
From this, it follows that
\bea
W^{ac} [D_{\mu},D_{\nu}] W_{cb} &=& 0
\eea
and this amounts to the following flatness condition for the field strength
\bea
\t F_{\mu\nu}{}_b{}^a + \t F_{\mu\nu}{}^a{}_b &=& 0.\label{fieldstrength}
\eea
Here, we defined
\bea
\t F_{\mu\nu}{}^a{}_b &=& W^{ac} W_{bd} \t F_{\mu\nu}{}_c{}^d \ .
\eea
By straightforward generalization, one can construct similar relations for monopole operators of higher charge.

A few remarks are in order.
First, for level $k=1$, we should in principle also be able to bring all matter fields into gauge singlets using $W_a$ and $W^a$ monopole operators. However, this does not give us any nice identity for the field strength. Instead, what we get is $F_{\mu\nu,a}{}^b W_b = 0$. However, we can not conclude from this any identity for $F_{\mu\nu}$ itself. It would be interesting to analyze how to use $W^a$ and $W_a$ to see supersymmetry and R-symmetry enhancement for level $k=1$. In our approach, we shall be using $W_{ab}$ and $W^{ab}$ for both $k=1$ and $k=2$.

Second, expanding $F_{\mu\nu} = F_{\mu\nu,\alpha} t^{\alpha}$ in the Lie algebra generators, one might be tempted to conclude from (\ref{fieldstrength}) that the Lie algebra generators are invariant under the similarity transformation induced by the monopole operator
\bea
(t^{\alpha})^b{}_a &=& -W_{ac} (t^{\alpha})^c{}_d W^{db}.\label{IDD}
\eea
This is not right because the gauge field strength cannot be varied independently of the monopole operator. Therefore (\ref{fieldstrength}) does not imply (\ref{IDD}). In fact, (\ref{IDD}) is not even gauge covariant since the generators do not transform under the gauge transformations whereas the monopole operators do transform in general. On the other hand, if we assume (\ref{IDD}), we find the BLG theory as the only solution for which $W_{ab} = \delta_{ab}$, the Kroenecker delta of the SO(4)$=$SU$_L(2)\times$SU$_R(2)$ gauge group (which is invariant, $\delta \delta_{ab} = \Lambda^c{}_a \delta_{cb} + \Lambda^c{}_b \delta_{ac} = \Lambda_{ba} + \Lambda_{ab} = 0$) and $(t^{\alpha})^a{}_b = (t^{\alpha})_{ab} = - (t^{\alpha})_{ba}$ are the antisymmetric generators of SO(4) gauge group. This is one of many indications that supersymmetry enhancement for the ABJM theory is highly nontrivial than one might naively extrapolate from the BLG theory.

Third, for a monopole operator $W_{a_1 \cdots a_k}$ of a representation involving $k$ Young tableaux boxes, the path independence implies that
\bea
D_\mu W_{a_1 \cdots a_k} = 0.
\eea
To see this, start with path integral of the gauge field. Upon taking singular gauge transformation of
`t Hooft loop operator and creating magnetic monopole of flux $\Phi$, the Chern-Simons action induces a Wilson line along `time' direction whose charge is $k \Phi$. This is the monopole operator whose representation is given by $k \Phi$ Young tableaux boxes. This is the direct generalization of the abelian case studied in section 3.
%%%%%%%%%%%%%%%%%%%%%%%%%%%%%%%%%%%%%%%%%%%%%%%%%%%%%%%%%%%%%%%%%%%%%%%%%%%%%%%%%%%%%%%%%%%%%%%%%%%%%%%
\section{Closure among ABJM and non-ABJM fields}

\subsection{closure relation and gauge condition}

As far as $\N=6$ and SU(4) symmetry variations (let us denote variations as $\delta$) are concerned, since ABJM fields and non-ABJM fields do not mix, we do not need to consider the quantities
\bea
{\Omega_\omega}^b{}_a &\equiv & W^{bc} \delta_\omega W_{ca},
\eea
which encodes variation of the monopole operator. On the other hand, when we explore possible $\N=8$ and SO(8) symmetry enhancement, we must consider these quantities since the ABJM and non-ABJM fields mix each other.
A priori, this indicates that we need to find explicit expression of ${\Omega^b}_a$.
This, however, turned out extremely difficult. Fortuitously, we never need the explicit expression, as we now explain below.

It is easy to see why ${\Omega^b}_a$ is needed when we mix the ABJM and non-ABJM fields. Let us assume that
\bea
\delta Z^A_a &=& \hat{\delta} Z^A_a \ ,
\eea
where $\hat{\delta}$ denotes any variation that does not involve ${\Omega^b}_a$ explicitly. We then get
\bea
\delta Z^{Aa} &=& \hat{\delta} Z^{Aa} - \Omega^a{}_b Z^{Ab}.
\eea
On the other hand, there is no good reason why ABJM fields should be treated any differently from non-ABJM fields. What we call ABJM and non-ABJM fields is really a matter of convention. Therefore, there is no reason we should not have ${\Omega^b}_a$ dependent terms in the variations of the ABJM fields. Let us therefore treat ABJM and non-ABJM fields on equal footing and take the general ansatz for the variations of the fields as
\bea
\delta Z^A_a &=& \hat{\delta} Z^A_a + \gamma \Omega^b{}_a Z^A_b\cr
\delta Z^{Aa} &=& \hat{\delta} Z^{Aa} + (\gamma-1) \Omega^a{}_b Z^{Ab}.
\eea
Here $\gamma$ could a priori be any real number. We then have
\bea
\delta (D_{\mu} Z^A_a) &=& \hat{\delta} (D_{\mu} Z^A_a) + \gamma \Omega^b{}_a D_{\mu} Z^A_b \ .
\eea
From the left-hand side, we get
\bea
\delta A_{\mu}^b{}_a &=& \hat{\delta} A_{\mu}{}^b{}_a + \gamma D_{\mu} \Omega^b{}_a \ .
\eea

Any symmetry variations should close among themselves. This requirement has an interesting consequence when it is applied to variations that mixes ABJM and non-ABJM fields. We get no restriction on $\gamma$ as long as we consider variations that do not mix ABJM and non-ABJM fields. Let us therefore consider SO(8) variations that mix these fields. We can also consider ${\cal N}=8$ variations but the steps are essentially the same. The variations take the form
\bea
\delta Z^A_a &=& \omega^{AB} Z_{Ba} + \gamma Z^A_b \Omega^b{}_a\cr
\delta Z^{Aa} &=& \omega^{AB} Z_B^a + (\gamma-1) \Omega^a{}_b Z^{Ab}\cr
\delta Z_{Aa} &=& -\omega_{AB} Z^B_a + (1-\gamma) Z_{Ab} \Omega^b{}_a \ . \label{variation}
\eea
More general variation may be considered such as $\hat{\delta} Z^A_a = \omega^A{}_B{}_a{}^b Z^{A}_b + \omega^{AB}{}_a{}^b Z_{Bb} + ...$ but the conclusion will anyway be the same. Since $Z^A_a$ and $Z_{Aa}$ transform the same under the gauge group and the second terms on the right hand side of the variations rotates gauge indices only, it motivates to have $\gamma = (1 - \gamma)$, viz.$\gamma =1/2$. We now show explicitly that this is indeed the necessary condition for the closure.

The closure among these variations reads
\bea
[\delta_{\eta},\delta_{\omega}] &=& \delta_{[\eta,\omega]} \ .
\eea
We get
\bea
[\delta_{\eta},\delta_{\omega}]Z^A_a &=& [\eta,\omega]^A{}_B Z^B_a \cr
&& + (1-\gamma) {\Omega_{\eta}}^b{}_a \omega^{AB} Z_{Bb} + \gamma {\Omega_{\omega}}^b{}_a \eta^{AB} Z_{Bb}\cr
&& - (1-\gamma) {\Omega_{\omega}}^b{}_a \eta^{AB} Z_{Bb} - \gamma {\Omega_{\eta}}^b{}_a \omega^{AB} Z_{Bb}\cr
&& + (\gamma^2 - \gamma) Z^A_b [\Omega_{\eta},\Omega_{\omega}]^b{}_a + \gamma Z^A_b \Omega_{[\eta,\omega]}{}^b{}_a \label{thisstep}
\eea
Here, we have used the variation
\bea
\delta_{\eta}{\Omega_{\omega}}^b{}_a &=& - \Omega_{\eta}{}^b{}_d \Omega_{\omega}{}^d{}_a + W^{bc} \delta_{\eta}\delta_{\omega} W_{ca} \ . \label{etaomega}
\eea
We also made the assumption that the variations close on the monopole operator
\bea
[\delta_{\eta},\delta_{\epsilon}] W_{ab} &=& \delta_{[\eta,\omega]} W_{ab}.
\eea
We now see that we can have the closure relation provided we set
\bea
\gamma &=& \frac{1}{2},
\eea
since in this case the mixed transformation terms cancel each other. The remaining terms read
\bea
[\delta_{\eta},\delta_{\omega}]Z^A_a &=& [\eta,\omega]^A{}_B Z^B_a \cr
&& + Z^A_b \(\frac{1}{2} \Omega_{[\eta,\omega]}{}^b{}_a - \frac{1}{4}[\Omega_{\eta},\Omega_{\omega}]^b{}_a \) \ .
\eea
Here, $\Omega$s form a closed algebra
\bea
[\Omega_\eta, \Omega_\omega] = \Omega_{[\eta, \omega]}
\eea
due to the fact that $\Omega$s are homomorphism of SO(8) to $\hat{\hat{G}}$.
Comparing with (\ref{variation}), we see that the closure relation is up to a gauge transformation:
\bea
[\delta_\eta, \delta_\omega] Z^A_a = \delta_{[\eta, \omega]} Z^A_a + \delta_{\rm gauge} Z^A_a
\eea
where the gauge parameter is given by $-{1 \over 4} \Omega_{[\eta, \omega]}$.

The result we found on $\gamma$ is very interesting. It means that we find a gauge variation with gauge parameter
\bea
\Lambda^b{}_a &=& \frac{1}{2} \Omega^b{}_a
\eea
induced from the SO(8) variations. This gauge variation can be off-set by making another gauge variation. This is the lucky circumstance that makes it possible to study variations that mix ABJM and non-ABJM fields without having to solve the tremendously difficult problem of finding an explicit expression for $\Omega^b{}_a$ or of the variation of the monopole operator itself.

%The factor of $\frac{1}{2}$ in the gauge parameter is confusing, but could perhaps %be attributed to the fact that the monopole operator $W^{ab}$ has two upper indices. %In the abelian case that would be a charge $2$ object. Hence in the abelian case we %would get
%\bea
%\Omega &=& 2\Lambda
%\eea
%if we take $\delta$ as a gauge variation.

Having seen that ${1 \over 2} \Omega$ is just a gauge parameter, we can just drop all $\Omega$-dependent terms from our variations from the outset.

%%%%%%%%%%%%%%%%%%%%%%%%%%%%%%%%%%%%%%%%%%%%%%%%%%%%%%%%%%%%%%%%%%%%%%%%%%%%%%%%%%%%%%%%%%%%%%%%%%%%%%%%
\subsection{combining gauge covariance with $\N=6$ supersymmetry}
We can use $\N=6$ supersymmetry to vary the identity (\ref{fieldstrength}) and get new identities. We can vary $\t F_{\mu\nu}$ either by varying its on-shell expression (\ref{gaugefield}), or we can compute the variation induced by variation of the gauge field as
\bea
\delta_{\epsilon} \t F_{\mu\nu} =  D_{\mu} \delta_{\epsilon} \t A_{\nu} - D_{\nu} \delta_{\epsilon} \t A_{\mu}.
\eea
Both computations give the same result when the fields are put on-shell.
%
%\footnote{Let us ignore 3-algebra terms and drop antihermitian conjugate term from our expressions, for simplicity. Then
%\bea \frac{1}{2}\epsilon^{\mu\nu\lambda} \delta_{\epsilon} F_{\mu\nu}{}^b{}_a &=& i\overline{\epsilon}_{AB} Z^A_c D^{\lambda} \psi^{Bd} - i\bar{\epsilon}_{AB} (D^{\lambda} Z^A_c)\psi^{Bd} - i \bar{\psi}^{Ad} \gamma^{\lambda}\gamma^{\mu} \epsilon_{AB} D_{\mu} Z^B_c + (3-alg) \eea
%Then we permute $\bar{\psi}^{Ad} \gamma^{\lambda} \gamma^{\mu} \epsilon_{AB} = 2 \bar{\epsilon}_{AB} \psi^{Ad} \eta^{\mu\lambda} - \bar{\epsilon}_{AB}\gamma^{\lambda}\gamma^{\mu}\psi^{Ad}$ and get \bea\frac{1}{2}\epsilon^{\mu\nu\lambda} \delta_{\epsilon} F_{\mu\nu}{}^b{}_a &=& i\bar{\epsilon}_{AB} D^{\lambda}(Z^A_c \psi^{Bd}) - i \bar{\epsilon}_{AB}\gamma^{\lambda}\gamma^{\mu}\psi^{Bd} D_{\mu} Z^A_c \eea
%On the other hand,
%\bea \epsilon^{\mu\nu\lambda} D_{\mu} \delta_{\epsilon} A_{\nu}{}^b{}_a &=& -i\bar{\epsilon}^{AB}\gamma^{\mu\lambda} D_{\mu}(\psi_{Ac}Z_B^d) f^{bc}{}_{da} + a.h.c. \eea
%We rewrite $\gamma^{\mu\lambda} D_{\mu} \psi_{Ac} = -\gamma^{\lambda}\gamma^{\mu} D_{\mu} \psi_{Ac} + D^{\lambda} \psi_{Ac}$ and use the fermionic equation of motion to write the first term in terms of 3-algebra terms. Then we have
%\bea \epsilon^{\mu\nu\lambda} D_{\mu} \delta_{\epsilon} A_{\nu}{}^b{}_a &=& i\bar{\epsilon}_{AB}D^{\lambda} (\psi^{Bd} Z^A_c) - i \bar{\epsilon}_{AB} \gamma^{\lambda} \gamma^{\mu} \psi^{Bd} D_{\mu} Z^A_c + (3-alg) \eea
%We see the two expressions match (3-algebra terms can be easily worked out and found to match as well).}
%
The latter approach is the quicker, and it gives the result
\bea
\delta_{\epsilon} \t F_{\mu\nu}{}^b{}_a &=& -i\overline{\epsilon}^{AB}\gamma_{\nu} D_{\mu}(\psi_{Ac}Z_B^d) f^{bc}{}_{da} + (\mbox{a.h.c}).
\eea
where (a.h.c) means that we should make the result antihermitian by adding the anti-hermitian conjugate term. Instead of computing the supersymmetry variation of $\t A_{\mu a}{}^b = \t A_{\mu}{}^d{}_c W^{bc}W_{da}$, we use the former approach and compute the variation of the on-shell field strength $\t F_{\mu\nu a}{}^b$
\bea
\t F_{\mu\nu a}{}^b &=& -\epsilon_{\mu\nu\lambda}\(Z^{Ad} D^{\lambda} Z_{Ac} - D^{\lambda} Z^{Ad} Z_{Ac}-i\overline{\psi}^{A}_c\gamma^{\lambda}\psi_{A}^d\)f^{bc}{}_{da}.
\eea
Then we can make a supersymmetry variation of the on-shell field strength. The result we get then is
\bea
\delta_{\epsilon} F_{\mu\nu a}{}^b &=& -i\overline{\epsilon}^{AB}\gamma_{\nu} D_{\mu}(\psi_{Ac} Z_{B}^d) f^{bc}{}_{da} + \mbox{(a.h.c.)} .
\eea

Now the $\N=6$ supersymmetry variation of the identity (\ref{fieldstrength}) reads
\bea
\overline{\epsilon}^{AB} \gamma_{[\nu} D_{\mu]} (\psi_{Ac} Z_B^d + \psi_A^d Z_{Bc}) f^{bc}{}_{da} + \mbox{(a.h.c.)} &=& 0.\label{id0}
\eea
$\overline{\epsilon}^{AB}$ and its conjugate are arbitrary, so we find the equations
\bea
\gamma_{[\nu} D_{\mu]} (\psi_{[Ac} Z_{B]}^d + \psi_{[A}^d Z_{B]c}) f^{bc}{}_{da} &=& 0.
\eea
From this equation it follows that
\bea
D_{\mu} (\psi_{[Ac} Z_{B]}^d + \psi_{[A}^d Z_{B]c}) f^{bc}{}_{da} &=& 0. \nonumber
\eea
To understand this we note that an equation $\gamma_{\nu}D_{\mu}\psi - \gamma_{\mu}D_{\nu}\psi =0$ implies $\gamma^{\mu}D_{\mu}\psi=0$ upon contracting by $\gamma^{\mu\nu}$. Second if we contract by $\gamma^{\mu}$ we find $-D_{\nu}\psi - \gamma^{\nu} (\gamma^{\mu} D_{\mu} \psi) = 0$. Hence $D_{\nu}\psi = 0$. The covariant derivative only acts on gauge indices, not on spinor indices. Since there is no independent covariantly constant spinor, we find six identities
\bea
(\psi_{[Ac} Z_{B]}^d + \psi_{[A}^d Z_{B]c}) f^{bc}{}_{da} &=& 0\label{main}
\eea
one for each choice of the antisymmetric indices $[AB]$. The right-hand side is zero since there is no non-trivial spinor of the same quantum number as the left-hand side.

It turns out (\ref{main}) is the supersymmetry variation of the identity:
\bea
(Z^A_c Z_{A}^d + Z^{Ad} Z_{Ac}) f^{bc}{}_{da} &=& 0.\label{niam}
\eea
Again we could have added a supersymmetric invariant to the right hand side, but there is no such an invariant which is also gauge covariant and has the same dimension.
To show this identity, take ${\cal N}=6$ supersymmetry transformation of (\ref{niam}):
\bea
0 &=& -i\overline{\epsilon}^{AB}\(\psi_{Bc} Z_A^d + \psi_B^d Z_{Ac}\)\cr
&&+i\overline{\epsilon}_{AB}\(Z^A_c\psi^{Bd} + Z^{Ad} \psi^B_c\)\cr
&&+ \frac{1}{2} \(Z^A_e Z_A^d + Z^{Ad} Z_{Ae}\) \(\Omega^e{}_c f^{bc}{}_{da} - \Omega^c{}_d f^{be}{}_{ca}\).
\eea

To get (\ref{main}) from this, we need to show that the third line vanishes. We note that $\Omega$ is a Lie algebra element, and hence we can pull out one 3-algebra structure constant from it as
\bea
\Omega^b{}_a &=& \omega^d{}_c f^{bc}{}_{da}
\eea
or we may directly use the fundamental identity (\ref{fundidentityinf}) $\delta f^{be}{}_{da} = 0$. Either way, we can rewrite the third line as
\bea
\frac{1}{2} \(Z^A_e Z_A^d + Z^{Ad} Z_{Ae}\) \(\Omega^b{}_c f^{ce}{}_{da} - \Omega^c{}_a f^{be}{}_{dc}\)
\eea
and this vanishes by the identity (\ref{niam}). This result is in concordance with the fact that $\Omega$-terms should play no important role in our equations.

We can generate identities involving three matter fields by making $\N=6$ supersymmetry variation of the identity
\bea
\t F_{ba} D^n Z_A^b + \t F_{ab} D^n Z_A^b &=& 0
\eea
where $D^n$ means any number $n$ derivatives. The non-trivial thing to be varied here is $\t F$. If we vary $D^n Z_A^b$ then we find a vanishing contribution by the identity $\t F_{ba} + \t F_{ab} = 0$. The new identities we generate this way read
\bea
(D^n Z_A) D_{\mu}(\psi_{[B} Z_{C]}) + D_{\mu}(\psi_{[B} Z_{C]}) (D^n Z_A)  &=& 0.
\eea
We want to conclude from this that we must have an identity
\bea
Z_A \psi_{[B} Z_{C]} + \psi_{[B} Z_{C]} Z_A  &=& 0
\eea
with no derivatives.

%%%%%%%%%%%%%%%%%%%%%%%%%%%%%%%%%%%%%%%%%%%%%%%%%%%%%%%%%%%%%%%%%%%%%%%%%%%%%%%%%%%%%%%%%%%%%%%%%%%%%%%%%
\section{$\N=8$ Supersymmetry}
We require any $\N=8$ supersymmetry variations be such they reproduce BLG variations for BLG gauge groups (that means SO(4) and such, for which $W_{ab} =  \delta_{ab}$ and $f^{bc}{}_{da} = f_{bcda}$ real and totally antisymmetric). We also require gauge covariance. We then find $\Omega$ terms that contribute a gauge variation with gauge parameter $\frac{1}{2}\Omega$. We off-set these by a supplementary gauge variation. Then we end up with the following ansatz for $\N=8$ supersymmetry variations (for levels $k=1,2$),
\bea
\delta Z_{Aa} &=& i\overline{\epsilon}_{AB} \psi^B_a - \overline{\epsilon} \psi_{Aa}\cr
\delta \psi_{Aa} &=& \gamma^{\mu} \epsilon_{AB} D_{\mu} Z^B_a + i \gamma^{\mu} \epsilon D_{\mu} Z_{Aa}\cr
&&+ \(\epsilon_{AB} Z^B_b Z^C_c Z_C^d + \epsilon_{BC} Z^B_b Z^C_c Z^d_A\) f^{bc}{}_{da}\cr
&&- i\epsilon Z_{Ab} Z^B_c Z_{B}^d f^{bc}{}_{da} + \frac{i}{3} \epsilon^* \epsilon_{ABCD} Z^B_b Z^C_c Z^{Dd} f^{bc}{}_{da},\cr
\delta \t A_{\mu}{}^b{}_a &=& \Big(-i\overline{\epsilon}^{AB} \gamma_{\mu} \psi_{Ac} Z_B^d + i \overline{\epsilon}_{AB}\gamma_{\mu}Z^A_c\psi^{Bd}\cr
&& + \overline{\epsilon} \gamma_{\mu} \psi_{Bc} Z^{Bd} + \overline{\epsilon}^* \gamma_{\mu} Z_{Bc} \psi^{Bd}\Big)f^{bc}{}_{da}, \label{theserules}
\eea
Much is surely getting fixed in these supersymmetry transformations by the requirement that it reproduces the BLG transformation rules in certain limits. We go through that argument in detail in Appendix using triality. Gauge covariance then dictates how to put the gauge 3-algebra indices, at least to a large extent. Still some ambiguities remain. We will see how that ambiguity is cured by having associated identities in section \ref{cured}.

It is also worth of noting that the supersymmetry transformations (\ref{theserules}) involve terms of baryon number $\Delta Q_{\rm B} = 0, \pm 1$. In M-theory, the baryon number is related to the Kaluza-Klein momentum around the M-theory circle. Upon dimensional reduction, there may be {\sl a priori} an infinite tower of fields carrying multiple Kaluza-Klein momentum. The fact that only fields with $\Delta Q_{\rm B} = 0, \pm 1$ and none with $\Delta Q_{\rm B} \ge 2$ appear implies that higher momentum modes are bound-states of these elementary modes.

%%%%%%%%%%%%%%%%%%%%%%%%%%%%%%%%%%%%%%%%%%%%%%%%%%%%%%%%%%%%%%%%%%%%%%%%%%%%%%%%%%%%%%%%%%%%%%%%%%%%%%%%%%%
\subsection{closing $\N=2$ supersymmetry}\label{cured}
The most general ansatz for the $\N=2$ supersymmetry variations such that they reduce to BLG variations for BLG gauge groups are given by a 3-parameter family (we denote the three parameters as $a$, $b$ and $d$ respectively):
\bea
\delta Z_{Aa} \ \ &=& -\overline{\epsilon} \psi_{Aa}\cr
\delta \psi_{Aa} \ &=& i \gamma^{\mu} \epsilon D_{\mu} Z_{Aa} - i \epsilon \(a Z_{Ab} Z^B_c Z_B^d + b Z^B_b Z_{Ac} Z_B^d + (1-a-b) Z_{Bb} Z^B_c Z_A^d\)f^{bc}{}_{da} \cr
&& + \frac{i}{3} \epsilon^* \epsilon_{ABCD} Z^B_b Z^C_c Z^{Dd} f^{bc}{}_{da},\cr
\delta \t A_{\mu}{}^b{}_a &=& \Big(-\overline{\epsilon} \gamma_{\mu} Z^A_c \psi_A^d - \overline{\epsilon}^* \gamma_{\mu} \psi^A_c Z_A^d\cr
&& + d \overline{\epsilon} \gamma_{\mu} \(Z^A_c \psi_A^d + Z^{Ad} \psi_{Ac}\) + d\overline{\epsilon}^* \gamma_{\mu} \(\psi^A_c Z_A^d + \psi^{Ad} Z_{Ac}\) \Big) f^{bc}{}_{da}
\eea
Eventually, we will see that all three parameters are traded for the three identities. At present, the only identity we can make use of, is identity in (\ref{niam}). We then find that the following variations
\bea
\delta Z_{Aa} \ \ &=& -\overline{\epsilon} \psi_{Aa}\cr
\delta \psi_{Aa} \ &=& i \gamma^{\mu} \epsilon D_{\mu} Z_{Aa} - i \epsilon Z_{Ab} (c Z^B_c Z_B^d - (1-c) Z^{Bd} Z_{Bc})f^{bc}{}_{da} + \frac{i}{3} \epsilon^* \epsilon_{ABCD} Z^B_b Z^C_c Z^{Dd} f^{bc}{}_{da},\cr
\delta \t A_{\mu}{}^b{}_a &=& - (\overline{\epsilon} \gamma_{\mu} (c Z^B_c \psi_B^d - (1-c) Z^{Bd} \psi_{Bc}) + \overline{\epsilon}^* \gamma_{\mu} (c \gamma_{\mu} \psi^B_c Z_B^d - (1-c) \psi^{Bd} Z_{Bc}) f^{bc}{}_{da}\label{two}
\eea
close on some equations of motion. More precisely, they close on the one parameter set of equations of motion
\bea
0 &=& \gamma^{\mu} D_{\mu} \psi_{Aa}\cr
 && + c \(2 Z_{Ab} Z^B_c \psi_B^d + \psi_{Ab} Z^B_c Z_B^d\) f^{bc}{}_{da}\cr
 && - (1-c) \(2 Z_{Ab} \psi_{Bc} Z^{Bd} + \psi_{Ab} Z_{Bc} Z^{Bd}\) f^{bc}{}_{da}\cr
 && + \frac{1}{3}\epsilon_{ABCD} (2 \psi^B_b Z^C_c Z^{Dd} + Z^B_b Z^C_c \psi^{Dd}) f^{bc}{}_{da} \ .
\eea
Of course, we can not really get different results since we use just one and the same supersymmetry variation, and the dependence on the parameter $c$ is fake, because we have the identity (\ref{niam}). So the equations of motion must not depend on the parameter $c$. This implies that
\bea
Z_{Ab} \(Z^B_c \psi_B^d + \psi_{Bc} Z^{Bd}\) f^{bc}{}_{da} &=& 0.\label{new}
\eea
Contracting this identity by $\psi^{Aa}$ we get
\bea
Z_{Ab} \psi^{Aa} \(Z^B_c \psi_B^d + \psi_{Bc} Z^{Bd}\) f^{bc}{}_{da} &=& 0
\eea
We then recall that
\bea
W^{ed}W_{fc} f^{bc}{}_{da} &=& W^{bg}W_{ah} f^{eh}{}_{gf}
\eea
which enable us to rewrite this identity as a perfect square
\bea
\left|\(Z^B_c \psi_B^d + \psi_{Bc} Z^{Bd}\) f^{bc}{}_{da}\right|^2 &=& 0
\eea
where $|M^b{}_a|^2 \equiv M^b{}_a M^{*b}{}_a$. Since each term in that sum is non-negative, we deduce the following set of identities
\bea
\(Z^B_c \psi_B^d + \psi_{Bc} Z^{Bd}\) f^{bc}{}_{da} &=& 0.
\eea

We must be have two more identities,
\bea
\(Z_{Ab} Z^B_c Z_B^d + Z_{Bb} Z^B_c Z_A^d\) f^{bc}{}_{da} &=& 0\label{nice}
\eea
and
\bea
\psi_{Bb} \(Z_{Ac} Z^{Bd} + Z^B_c Z_A^d\) f^{bc}{}_{da} &=& 0\label{leftover}
\eea
in order to be able to close $\N=2$ variations among themselves. These identities can not be derived without using $\N=6$ supersymmetry though, and we will return to the derivation of these identities in the next section. For the time being let us note that the identity (\ref{nice}) is required in order to make the ansatz for the $\N=2$ supersymmetry variations free from ambiguities despite there are three free parameters. To each of these parameters there will be an associated identity and so there will be no ambiguities.

From the identity (\ref{nice}) we can derive another identity
\bea
\(\psi^C_b Z_{Bc} Z^{Bd} - Z_{Bb} Z^B_c \psi^{Cd}\)f^{bc}{}_{da} &=& 0\label{what}
\eea
To see this we make an $\N=5$ supersymmetry variation of (\ref{nice}). That gives us
\bea
0 &=& \Sigma^M_{AD} \(\psi^D_b Z^B_c Z_B^d - Z^B_b Z_{Bc} \psi^{Dd}\)\cr
&& + \Sigma^{M,BD} \(Z_{Ab} \psi_{[Dc} Z_{B]}^d + Z_{[Bb} \psi_{D]c} Z_A^d\)\cr
&& + \Sigma^M_{BD} \(Z_{Ab} Z^{[Bc} \psi^{D]d} + \psi^{[Db} Z^{B]c} Z_A^d\).
\eea
Then the identity $Z_{Bb} Z^D_c \psi^{Bd} + \psi^B_b Z^D_c Z_B^d = 0$ follows from the identity Eq (\ref{main}). Hence, we are left with the identity in (\ref{what}). Consequently, if we can establish (\ref{what}) then we can also be sure that (\ref{nice}) holds.

Given these identities, we find the following closure relations for the $\N=2$ supersymmetry variations,
\bea
[\delta^{(2)}_{\eta},\delta^{(2)}_{\epsilon}] Z_{Aa} \ &=& -2i \overline{\epsilon}^X \gamma^{\mu} \eta^X D_{\mu} Z_{Aa} + \t \Lambda^{(22)}{}^b{}_a Z_{Ab}\cr
[\delta^{(2)}_{\eta},\delta^{(2)}_{\epsilon}] \psi_{Aa} &=& -2i \overline{\epsilon}^X \gamma^{\mu} \eta^X D_{\mu} \psi_{Aa} + \t \Lambda^{(22)}{}^b{}_a \psi_{Aa}\cr
&& + i \overline{\epsilon}^X \gamma^{\lambda} \eta^X \gamma_{\lambda} E^{(22)}_{Aa} -2 \overline{\epsilon}^{[8} \eta^{7]} E^{(22)}_{Aa},\cr
[\delta^{(2)}_{\eta},\delta^{(2)}_{\epsilon}] \t A_{\mu}{}^b{}_a &=& -2i \overline{\epsilon}^X \gamma^{\mu} \eta^X \t F_{\nu\mu}{}^b{}_a - D_{\mu} \t \Lambda^{(22)}{}^b{}_a
\eea
with gauge parameter
\bea
\t \Lambda^{(22)}{}^b{}_a &=& 4 \overline{\epsilon}^{[8} \eta^{7]} Z^B_c Z_B^d f^{bc}{}_{da}
\eea
and we have closure on the ABJM equations of motion.

%%%%%%%%%%%%%%%%%%%%%%%%%%%%%%%%%%%%%%%%%%%%%%%%%%%%%%%%%%%%%%%%%%%%%%%%%%%%%%%%%%%%%%%%%%%%%%%%%%%%%%%%%%%
\subsection{commuting $\N=6$ and $\N=2$ supersymmetries}
As all identities are of essentially the same form we find it more transparent if we introduce a short-hand notation. We write $XY$ as short for $X_c Y^d f^{bc}{}_{da}$ and $XYZ$ as short for $X_b Y_c Z^d f^{bc}{}_{da}$. We write $U[XYZ]$ as short for $U^a X_b Y_c Z^d f^{bc}{}_{da}$.

Making an $\N=6$ supersymmetry variation of identity (\ref{new}) we obtain three new identities\footnote{To understand how we can get three new identies instead of just one, we note that an equation of the form
\bea
\gamma^{\mu} \epsilon_M V_{\mu} + \epsilon_M U &=& 0
\eea
with $\epsilon_M$ arbitrary, implies that $U=0$ and $V_{\mu}=0$ separately.},
\bea
Z^{[A} D_{\mu} Z^{B]} + (D_{\mu} Z^{[B}) Z^{A]} &=& 0\\
Z^{[A}[Z^{B]} Z^C Z_C] + [Z^{[B} Z^C Z_C] Z^{A]} + Z^C[Z^A Z^B Z_C] + [Z^A Z^B Z_C] Z^C &=& 0\label{four}\\
\overline{\psi}_{[A} \gamma^{\mu} \psi_{B]} &=& 0.
\eea
To be able to close supersymmetry and show SO(8) invariance, we must have two more identities. These are
\bea
Z_{A} Z_{B} Z_C - Z_{C} Z_{[A} Z_{B]} &=& 0,\cr
Z_{A} Z_{B} Z^{C} - Z^C Z_{[A} Z_{B]} &=& 0.\label{cyc}
\eea
We derive these identities as follows. Contracting the first equation by the totally independent spinor $\psi^{Ca}$, we see that the result vanishes by identities (\ref{main}), (\ref{what}). As an unnecessary extra check we can also contract the left-hand side by $Z^{Ca}$ and again get zero by identity (\ref{nice}). Now we have more than shown that this identity holds. The second identity is proved the same way, by contracting by $\psi_{C}^a$.

Let us make an $\N=6$ supersymmetry variation of the first identity. Expanding $\Sigma^{M,AB} \Sigma^N_{CD}$ using Fierz relations in appendix, we find the supersymmetry variation gives just one single set of identities,
\bea
\psi^C Z_{[A} Z_{B]} - Z_{A} Z_{B} \psi^{C} &=& 0.\label{ett}
\eea

Using the same method as above, but applied to mixed supersymmetry variations\footnote{In practive this means we compute $\delta^{(6)}_{\epsilon} (a Z_A Z_B Z^B + b Z^B Z_A Z_B + (1-a-b) Z_B Z^B Z_A)$ and require the result be independent of $a$ and $b$.}, we generate the following new identities
\bea
Z_{A} \psi^{[D} Z^{B]} - \psi^{[D} Z^{B]} Z_A &=& 0\label{tva}\\
Z_{A} Z_{[B} \psi_{D]} + \psi_{[D} Z_{B]} Z_A &=& 0
\eea

Using this we can now also derive the identity (\ref{nice}) that we had left-over from the previous section. We start by specializing identity (\ref{ett}) to the identity (we also complex conjugate)
\bea
-Z_B Z^{[B} \psi^{A]} + \psi^{[B} Z^{A]} Z_B &=& 0.
\eea
Expanding this out, we have
\bea
\(Z_B Z^A \psi^B + \psi^B Z^A Z_B\) - \(Z_B Z^B \psi^A - \psi^A Z_B Z^B\) &=& 0.
\eea
We may rewrite the first parentesis as $Z^A \(Z_B \psi^B + \psi^B Z_B\)$. We now see that this vanishes by an identity. Then the second parentesis must also vanish and we obtain the identity (\ref{what}), which in turn follows from (\ref{nice}) by a supersymmetry variation.

Finally we derive (\ref{leftover}) by specializing (\ref{tva}) to the identity
\bea
\psi_{[B} Z_{A]} Z^B - Z^B \psi_{[B} Z_{A]} &=& 0.
\eea
We expand out this as
\bea
\psi_B \(Z_A Z^B + Z^B Z_A\) - \psi_A \(Z_B Z^B + Z^B Z_B\) &=& 0
\eea
The second parentesis vanishes by a by now familiar identity, and (\ref{leftover}) follows.

We have now completed the derivation of all identities we need to close $\N=8$ supersymmetry on-shell.

Commuting an $\N=2$ and an $\N=6$ variation, given the above identities, we get
\bea
([\delta^{(2)}_{\eta},\delta^{(6)}_{\epsilon}] + [\delta^{(6)}_{\eta},\delta^{(2)}_{\epsilon}])  Z_{Aa} &=& \t \Lambda^b{}_a Z_{Ab},\cr
([\delta^{(2)}_{\eta},\delta^{(6)}_{\epsilon}] + [\delta^{(6)}_{\eta},\delta^{(2)}_{\epsilon}]) \psi_{Aa} &=& \t \Lambda^b{}_a \psi_{Ab} - (\epsilon_{AB} \eta - \eta_{AB} \epsilon) E_a^B,\cr
([\delta^{(2)}_{\eta},\delta^{(6)}_{\epsilon}] + [\delta^{(6)}_{\eta},\delta^{(2)}_{\epsilon}]) \t A_{\mu}{}^b{}_a &=& - D_{\mu} \t \Lambda^b{}_a
\eea
with gauge parameter
\bea
\t \Lambda^{(62)}{}^b{}_a &=& \(\overline{\epsilon}\eta_{AB} Z^A_c Z^{Bd} + \overline{\epsilon}^* \eta^{AB} Z_{Ac} Z_B^d\) f^{bc}{}_{da} - (\epsilon \leftrightarrow \eta)
\eea
and we have closure on the ABJM fermionic equation of motion.
% $E_{Aa} = 0$.

%%%%%%%%%%%%%%%%%%%%%%%%%%%%%%%%%%%%%%%%%%%%%%%%%%%%%%%%%%%%%%%%%%%%%%%%%%%%%%%%%%%%%%%%%%%%%%%%
\section{Manifestly SO(8) invariant ABJM scalar potential}\label{sextic}
In the previous section, we asked how ${\cal N}=6$ supersymmetry of the ABJM theory can be enhanced to ${\cal N}=8$ supersymmetry at $k=1, 2$. In this section, we shall ask analogous question: how SU(4) R-symmetry of the ABJM theory can be enhanced to SO(8) R-symmetry. Once again, we find that the enhancement takes place at $k=1,2$, for which the symmetric rank-2 monopole operator becomes local and plays the essential role that allows rotation between ${\bf 4}$ and $\overline{\bf 4}$ in a manner compatible with gauge covariance.

For concreteness, we shall focus on the ABJM sextet potential. The consideration extends to the ABJM Yukawa interactions in precisely the same way as the sextet potential. First of all, the ABJM sextet potential is expressible mostly compactly using the 3-bracket formulation. It takes the form
\bea
V_{\rm ABJM} = \frac{2}{3}\(\|[Z^A, Z^B; Z^C]\|^2 - \frac{1}{2} \|[Z^A, Z^B; Z^A]\|^2\) \ ,
\eea
where
\bea
\|X\|^2 &:=& \<X,X\>
\eea
and $SU(4)$ indices $A, B, C, \cdots$ are contracted. We note that in this notation all SU(4) indices are up-stairs despite some of them are being contracted. As before, in this notation, any time an SU(4) index is found down-stairs, that will correspond to a non-ABJM field -- a elementary field with the symmetric rank-2 monopole operator attached.

For the sake of completeness, let us list a few equivalent ways of expressing the ABJM sextet potential. We have the following alternative expressions
\bea
V &=& \frac{2}{3} \|[Z^A,Z^B;Z^C] + \alpha [Z^D,Z^{[A};Z^D] \delta^{B]}_C\|^2\cr
V &=& \frac{2}{3}\(f^{ab}{}_{gh} f^{ch}{}_{ef} - \frac{1}{2} f^{ab}{}_{eh} f^{ch}{}_{gf}\) Z^A_a Z_A^e Z^B_b Z_B^f Z^C_c Z_C^g
\eea
in the 3-algebra formulation. Here, we can choose either $\alpha = \frac{1}{2}$ or $\alpha = \frac{1}{6}$.
% 12 \alpha^2 - \alpha + 1 = 0 by brute-force expansion.
In the matrix realization of the 3-algebra, we find the potential expressed as
\bea
V &=& - \frac{1}{3}\tr\Big(Z^AZ_A Z^BZ_B Z^C Z_C + Z_A Z^A Z_B Z^B Z_C Z^C\cr
&& \qquad + 4 \ Z^A Z_C Z^B Z_A Z^C Z_B - 6 Z^A Z_C Z^B Z_B Z^C Z_A\Big)
\eea
and as it should, this vanishes when the matrices are commuting.

We now show that the ABJM potential can be written in the manifestly SO(8) invariant form of hermitian BLG theory:
\bea
V_{\rm BLG}&=& \frac{1}{12}\|[Z^{\alpha},Z^{\beta};Z^{\gamma}]\|^2.
\eea
Here, $Z^{\alpha}$ are real-valued SO(8) spinors and $Z^{\alpha}$ is not distinguished from $Z_{\alpha}$. Expanding the fields as $Z^{\alpha}=(Z^A,Z_A)$, viz. $Z^{\alpha}_a = (Z^A_a, Z_{Aa})$ where $Z_A$ has a monopole operator attached, we can express the BLG scalar potential in the form
\bea
V_{\rm BLG} &=& \frac{1}{6}\(\|[Z^A,Z^B;Z_C]\|^2 + \|[Z^A,Z^B;Z^C]\|^2 + 2\|[Z^A,Z_B;Z^C]\|^2\).
\label{blgV2}
\eea

We will now prove that the BLG scalar potential in the above form is identical to the ABJM potential {\sl once} the algebraic identities derived in the previous section are taken into account. First, we use the identities (\ref{cyc}) and put the BLG potential (\ref{blgV2}) further in the form:
\bea
V_{\rm BLG} &=& \frac{1}{6}\(\<[Z^A,Z^B;Z_C],[Z^A,Z^B;Z_C]\> + 3\<[Z^A,Z^B;Z^C][Z^A,Z^B;Z^C]\>\).
\eea
Next, we use the fundamental identity (\ref{fi}) together with the trace invariance condition (\ref{trinv}) and derive the following trace identity:
\bea
\<[X,Y;Z],[U,V;W]\> &=& \<[X,W;V],[U,Z;Y]\>\cr
&& - \<[Y,W;V],[U,Z;X]\>\cr
&& + \<[X,Y;U],[Z,V;W]\> \ .   \label{traceidentity}
\eea
Notice that the fundamental identity (\ref{fi}) is essentially the same algebraic structure as in the BLG theory, the only difference being that the ABJM 3-algebra is a refined version of the BLG 3-algebra where care must be taken for the way the generators are ordered inside the 3-product. Notice also that the condition (\ref{trinv}) is the same trace invariance condition as in the BLG theory, the only difference being that care must be taken for the ordering of elements. By applying (\ref{traceidentity}), we derive another identity:
\bea
\<[Z^A,Z^B;Z_C],[Z^A,Z^B;Z_C]\> &=& \<[Z^A,Z_C;Z^B],[Z^A,Z_C;Z^B]\>\cr
&& - \<[Z^B,Z_C;Z^B],[Z^A,Z_C;Z^A]\>\cr
&& + \<[Z^A,Z^B;Z^A],[Z_C,Z^B;Z_C]\> \ . \label{traceidentity}
\eea
Now, we rewrite the last term as
\bea
[Z_C,Z^B;Z_C] &=& [-Z^C,Z^B;Z^C]
\eea
using the identity (\ref{cyc}) and the second term as
\bea
[Z^B,Z_C;Z^B] &=& [Z_B,Z_C;Z_B]
\eea
again using (\ref{cyc}).  Using this, we can write the trace identity (\ref{traceidentity}) in the form
\bea
\<[Z^A,Z^B;Z_C],[Z^A,Z^B;Z_C]\> &=& \<[Z^A,Z_C;Z^B],[Z^A,Z_C;Z^B]\>\cr
&& - 2\<[Z^B,Z_C;Z^B],[Z^A,Z_C;Z^A]\> \ .
\eea
Substituting this expression into the hermitian BLG potential, we find that this becomes precisely the same as the ABJM potential. This establishes the sought-for SO(8) invariance of the ABJM scalar potential.

It is interesting to observe that, despite the 3-brackets are a priori antisymmetric only in its first two entries, these 3-brackets are totally antisymmetric in all its entries. That is,
\bea
V_{\rm BLG} = {1 \over 12} \vert\!\vert \ [ \ Z^{[\alpha}, Z^\beta; Z^{\gamma ]} \ ] \ \vert\!\vert \ .
\eea
We now show that this remarkable bonus property follows again from the algebraic identities we derived in the previous sections. First, we consider the first term in the ABJM potential and just apply the identity (\ref{cyc}), which in terms of 3-brackets reads
\bea
[Z^A,Z^B;Z^C] &=& \frac{1}{2}\([Z_C,Z^A;Z_B] - [Z_C,Z^B;Z_A]\).
\eea
Again, notice that the right hand side involves two non-ABJM fields, viz. two monopole operators. We then get
\bea
\<[Z^A, Z^B; Z^C],[Z^A, Z^B; Z^C]\> &=& \<[Z^A, Z^B; Z^C],[Z_C,Z^A;Z_B]\>\cr
&=& -\<[Z^A, Z^B; Z^C],[Z^A, Z_C; Z_B]\>
\eea
and we can continue from here as
\bea
-\<[Z^A, Z^B; Z^C],[Z^A, Z_C; Z_B]\> &=& \<[Z^B, Z^A; Z^C],[Z^A, Z_C; Z_B]\>\cr
&=& \<[Z^A,Z_C;Z_B],[Z^A,Z_C;Z_B]\>.
\eea
Of course it is not true that
\bea
[Z^A,Z^B;Z^C] &=& -[Z^A,Z_C;Z_B]
\eea
For this to be true we must contract by something antisymmetric in $BC$. However, there is no way to really tell whether this is the case or not by just looking at the first term -- this term behaves in all respects just as if the 3-bracket had been totally antisymmetric.

For the second term we have by identities
\bea
[Z^A, Z^B; Z^A] &=& -[Z^A, Z_A;Z_B].
\eea
Hence, the terms are totally antisymmetric. This completes the proof.

\vskip40pt
\subsection*{Acknowledgements} We thank Dongsu Bak, David J. Gross, Jeong-Hyuck Park, C. Socchiciu
and Takao Suyama for useful discussions and comments, and Martin Cederwall, Neil Lambert, Hai Lin and Anastasios Petkou for several correspondences. SJR thanks the hospitality of the Kavli Institute for Theoretical Physics during the course of this work. This work was supported in part by the National Research Foundation grants SRC-R11-2005-021, 2005-009-3843, 2009-008-0372, 2010-220-C00003, EU-FP Marie Curie Training Program (KICOS-2009-06318) and in part by the U.S. National Science Foundation under Grant No. PHY05-51164 at KITP (SJR).

\vskip20pt
Note added: After this paper was submitted on ArXiv, there appeared further works exploiting the supersymmetry enhancement for ABJM theory with higher-rank gauge group $U(N_1)_k \times U(N_2)_{-k}$ at $k=1, 2$ \cite{furtherworks}.

\newpage
\appendix
\section{SO(8) gamma matrices}
Here $\Gamma^I$ are SO(8) gamma matrices in the Weyl basis
\bea
\Gamma^I &=& \(\begin{array}{cc}
0 & \Gamma^{I\alpha\dot{\beta}}\\
\Gamma^I_{\dot{\alpha}\beta} & 0
\end{array}\)
\eea
They can be chosen to have real components and are then antisymmetric
\bea
(\Gamma^I)^T &=& -\Gamma^I.
\eea
The charge conjugation matrix is then
\bea
\Omega &=& \(\begin{array}{cc}
\delta_{\alpha\beta} & 0\\
0 & \delta^{\dot{\alpha}\dot{\beta}}
\end{array}\)
\eea
and its inverse is
\bea
\Omega^{-1} &=& \(\begin{array}{cc}
\delta^{\alpha\beta} & 0\\
0 & \delta_{\dot{\alpha}\dot{\beta}}
\end{array}\)
\eea
Since invariant tensors with two equal indices (that is $\delta_{IJ}$, $\delta_{\alpha\beta}$ and $\delta_{\dot{\alpha}\dot{\beta}}$) in SO(8) are thus identity matrices, we can put all SO(8) indices downstairs. We define the chirality matrix
\bea
\Gamma &=& \Gamma^{12...8}
\eea
These gamma matrices have properties
\bea
\Gamma^2 &=& 1\cr
\{\Gamma,\Gamma^I\} &=& 0\cr
\Gamma^T &=& \Gamma\cr
(\Gamma^I)^T &=& -\Gamma^I\cr
(\Gamma^{IJ})^T &=& -\Gamma^{IJ}\cr
(\Gamma^{IJK})^T &=& \Gamma^{IJK}\cr
(\Gamma^{IJKL})^T &=& \Gamma^{IJKL}
\eea
and duality
\bea
\Gamma^{I_1...I_m} &=& \frac{1}{(8-m)!} \epsilon^{I_1...I_m I_{m+1} ... I_8} \Gamma \Gamma^{I_8... I_{m+1}}\cr
\Gamma \Gamma^{I_8...I_m} &=& \frac{1}{(m-1)!}\epsilon^{I_1...I_m I_{m+1}...I_8} \Gamma^{I_1...I_{m-1}}
\eea
Defining $\overline{\eta} = \epsilon^{\dag}$, we find the the Fierz identity
\bea
16 \epsilon\overline{\eta} &=&-(\overline{\eta}\epsilon)-(\overline{\eta}\Gamma\epsilon)\Gamma\cr
&&-(\overline{\eta}\Gamma^I\epsilon)\Gamma^I+(\overline{\eta}\Gamma\Gamma^I\epsilon)\Gamma\Gamma^I\cr
&&-\frac{1}{2}(\overline{\eta}\Gamma^{IJ}\epsilon)\Gamma^{IJ}+\frac{1}{2}(\overline{\eta}\Gamma\Gamma^{IJ}\epsilon)\Gamma\Gamma^{IJ}\cr
&&+\frac{1}{6}\((\overline{\eta}\Gamma^{IJK}\epsilon)\Gamma^{IJK}-(\overline{\eta}\Gamma\Gamma^{IJK}\epsilon)\Gamma\Gamma^{IJK}\)\cr
&&-\frac{1}{24}(\overline{\eta}\Gamma^{IJKL}\epsilon)\Gamma^{IJKL}
\eea
For chiral spinors
\bea
\Gamma \epsilon &=& \epsilon\cr
\Gamma \eta &=& \eta
\eea
we have
\bea
\overline{\eta} \Gamma^{I_1...I_{odd}} \epsilon = 0
\eea
and get the Fierz identity
\bea
\epsilon \overline{\eta} &=& \frac{1}{16}\[-\overline{\eta}\epsilon + \frac{1}{2} \overline{\eta} \Gamma_{IJ} \epsilon \Gamma_{IJ} - \frac{1}{24} \overline{\eta}\Gamma_{IJKL}\epsilon \Gamma_{IJKL}\] (1+\Gamma).
\eea
and consequently
\bea
16 \(\epsilon \overline{\eta} - \eta\overline{\epsilon}\) &=& \overline{\eta} \Gamma_{IJ} \epsilon \Gamma_{IJ} \frac{1+\Gamma}{2}.
\eea

%%%%%%%%%%%%%%%%%%%%%%%%%%%%%%%%%%%%%%%%%%%%%%%%%%%%%%%%%%%%%%%%%%%%%%%%%%%%%%%%%%%%%%%%%%%%%
\section{SO(1,2) gamma matrices}
We let $\gamma^{\mu}$ denote gamma matrices and $c$ charge conjugation. These have properties
\bea
c^T &=& -c\cr
(\gamma^{\mu})^T &=& -c\gamma^{\mu} c^{-1}
\eea
We have the Fierz identity
\bea
\epsilon\overline{\eta} &=& -\frac{1}{2}\(\overline{\eta}\epsilon+(\overline{\eta}\gamma^{\mu}\epsilon)\gamma_{\mu}\).
\eea
An explicit realization is
\bea
\gamma^0 = \(\begin{array}{cc}
0 & 1\\
-1 & 0
\end{array}\),
\gamma^1 = \(\begin{array}{cc}
0 & 1\\
1 & 0
\end{array}\),
\gamma^2 = \(\begin{array}{cc}
1 & 0\\
0 & -1
\end{array}\)
\eea
and
\bea
c &=& \(\begin{array}{cc}
0 & -1\\
1 & 0
\end{array}\)
\eea
Since also
\bea
(\gamma_{\mu})^{\dag} &=& \gamma_0 \gamma_{\mu} \gamma_0\cr
(\gamma_{\mu})^{T} &=& -c\gamma_{\mu}c^{-1}
\eea
and we understand that the choice
\bea
c = \gamma_0
\eea
amounts to gamma matrices with real components, for instance we could take them as specified explicitly above.

In such a basis, Majorana spinors also have real components since the majorana condition
\bea
\overline{\psi} = \psi^T c
\eea
amounts to the condition
\bea
\psi^{\dag} = \psi^T
\eea
if we define $\overline{\psi} =\psi^{\dag} \gamma_0$.

%%%%%%%%%%%%%%%%%%%%%%%%%%%%%%%%%%%%%%%%%%%%%%%%%%%%%%%%%%%%%%%%%%%%%%%%%%%%%%%%%%%%%%%%%%%%%%%%%%%%%%
\section{Reducing SO(8) to SU(4)$\times$ U(1)}\label{gamma}
To reduce BLG theory to ABJM theory we want to reduce the symmetry as
\bea
\mbox{SO(8)} \rightarrow \mbox{SO(6)}\times \mbox{SO(2)} = \mbox{SU(2)} \times \mbox{U(1)} \ .
\eea
We represent the SO(8) gamma matrices
\bea
\Gamma^I = (\Gamma^M, \Gamma^7, \Gamma^8)
\eea
where
\bea
\Gamma^M &=& \Sigma^M \otimes \sigma^1\cr
\Gamma^7 &=& 1 \otimes \sigma^2\cr
\Gamma^8 &=& \Sigma \otimes \sigma^1
\eea
and
\bea
\sigma^1 &=& \(\begin{array}{cc}
0 & 1\\
1 & 0
\end{array}\),\cr
\sigma^2 &=& \(\begin{array}{cc}
0 & -i\\
i & 0
\end{array}\)
\eea
Here $\Sigma^M$ are hermitian SO(6) gamma matrices that we represent as
\bea
\Sigma^M &=& \(\begin{array}{cc}
0 & \Sigma^{M,AB}\\
\Sigma^M_{AB} & 0
\end{array}\)
\eea
where $A$ is Weyl index of SO(6), its chirality being distinguished by the placing up and down respectively. Hermiticity amounts to the condition
\bea
\Sigma^{*M,AB} &=& -\Sigma^M_{AB}.
\eea
We also define
\bea
\Sigma &=& \(\begin{array}{cc}
\delta^A{}_B & 0\\
0 & -\delta_A{}^B
\end{array}\).
\eea
We use index notation as follows. The spinor and co-spinor are decomposed as
\bea
\xi_{\alpha} &=& \(\begin{array}{c}
\xi^A\\
\xi_A
\end{array}\)\cr
\xi_{\dot{\alpha}} &=& \(\begin{array}{c}
\xi^A\\
-\xi_A
\end{array}\).
\eea
Accordingly, matrices (linear maps on the space of these vectors) are represented as
\bea
M_{\alpha\dot{\beta}} &=& \(\begin{array}{cc}
M^A{}_B & M^{AB}\\
M_{AB} & M_A{}^B
\end{array}\),\cr
M_{\dot{\alpha}\beta} &=& \(\begin{array}{cc}
M^A{}_B & M^{AB}\\
M_{AB} & M_A{}^B
\end{array}\),\cr
M_{\alpha\beta} &=& \(\begin{array}{cc}
M^A{}_B & M^{AB}\\
M_{AB} & M_A{}^B
\end{array}\),\cr
M_{\dot{\alpha}\dot{\beta}} &=& \(\begin{array}{cc}
M^A{}_B & M^{AB}\\
M_{AB} & M_A{}^B
\end{array}\)
\eea
and these in turn sit in an SO(8) matrix
\bea
\(\begin{array}{cc}
M_{\alpha\beta} & M_{\alpha\dot{\beta}}\\
M_{\dot{\alpha}\beta} & M_{\dot{\alpha}\dot{\beta}}
\end{array}\)
\eea
that maps a spinor $(\xi_{\alpha},\xi_{\dot{\alpha}})^T$ into a new spinor with the same spinor index structure.

For the reduction we also need
\bea
\Gamma_{IJ} &=& \(\Gamma_{MN}, \Gamma_{M7}, \Gamma_{M8}, \Gamma_{78}\)\cr
&=& \(\Sigma_{MN}\otimes 1,\Sigma_M \otimes i\sigma^3, \Sigma_M \Sigma \otimes 1, -\Sigma \otimes i\sigma^3\)
\eea

We define the hermitian SO(8) chirality matrix as
\bea
\Gamma &=& i \Gamma^{1...8} = 1 \otimes \sigma^3.
\eea

It is conventient to define supersymmetry parameter
\bea
\epsilon^{AB} = \epsilon^M \Sigma^{M,AB}
\eea
where $\epsilon^M$ is a real component spinor. This will have the property
\bea
(\epsilon^{AB})^* &=& -\epsilon_{AB}\cr
(\epsilon_{AB})^* &=& -\epsilon^{AB}
\eea

We have that
\bea
\Sigma^M_{AB} = \frac{1}{2}\epsilon_{ABCD}\Sigma^{M,CD} = \Sigma^{*M,BA}
\eea
and
\bea
\Sigma^M_{AB} \Sigma^{M,CD} &=& -4\delta_{AB}^{CD}\cr
\Sigma^M_{AB} \Sigma^M_{CD} &=& -2\epsilon_{ABCD}.
\eea

\vskip0.3cm
%%%%%%%%%%%%%%%%%%%%%%%%%%%%%%%%%%%%%%%%%%%%%%%%%%%%%%%%%%%%%%%%%%%%%%%%%%%%%%%%%%%%%%%%%%%%%%%%%%%
\section{Some more useful relations}
The ${\cal{N}}=8$ Fierz identity is
\bea
\epsilon_I \overline{\eta}_J - \eta_I \overline{\epsilon}_J &=& -\overline{\epsilon}_{[I}\eta_{J]}+\overline{\epsilon}_{(I}\gamma^{\mu}\eta_{J)}\gamma_{\mu},
\eea

\subsection{${\cal{N}}=6$}
Fierz identities read
\bea
\Sigma^M_{AB}\Sigma^N_{CD} &=& -(\Sigma^{MN})_{[A}{}^E \epsilon_{|E|B]CD}-\frac{1}{3}\delta^{MN}\epsilon_{ABCD}\cr
\Sigma^M_{AB}\Sigma^{N,CD} &=& -2(\Sigma^{MN})_{[A}{}^{[C}\delta_{B]}^{D]} - \frac{2}{3} \delta^{MN} \delta_{AB}^{CD}.
\eea

\subsection{${\cal{N}}=2$}
Fierz identities read
\bea
\epsilon \overline{\eta} - \eta \overline{\epsilon} &=& (\overline{\epsilon}_X \gamma^{\nu} \eta_X) \gamma_{\nu} - 2i \overline{\epsilon}_{[8} \eta_{7]}\cr
\epsilon^* \overline{\eta}^* - \eta^* \overline{\epsilon}^* &=& (\overline{\epsilon}_X \gamma^{\nu} \eta_X) \gamma_{\nu} + 2i \overline{\epsilon}_{[8} \eta_{7]}\cr
\epsilon \overline{\eta}^* - \eta \overline{\epsilon}^* &=& \(\overline{\epsilon}_7 \gamma^{\nu} \eta_7-\overline{\epsilon}_8 \gamma^{\nu} \eta_8 + 2i \overline{\epsilon}_{(8}\gamma^{\nu}\eta_{7)}\)\gamma_{\nu}
\eea
and then we have
\bea
\overline{\epsilon} \gamma^{\mu} \eta - \overline{\eta} \gamma^{\mu} \epsilon&=& 2\overline{\epsilon}^X \gamma^{\mu} \eta^X\cr
\overline{\epsilon} \eta - \overline{\eta} \epsilon &=& -4i\overline{\epsilon}^{[8}\eta^{7]}\cr
\overline{\epsilon}^* \eta - \overline{\eta}^* \epsilon &=& 0.
\eea

\subsection{$\N=8$}
Fierz identities are those for $\N=6$ and $\N=2$ plus the mixed ones,
\bea
\epsilon_M \overline{\eta} - \eta_M \overline{\epsilon} &=& \frac{1}{2}\(-\overline{\epsilon}_M \eta^* + \overline{\epsilon}_M\gamma^{\mu} \eta^* \gamma_{\mu}\) - (\epsilon \leftrightarrow \eta)\cr
\epsilon \overline{\eta}_M - \eta \overline{\epsilon}_M &=& \frac{1}{2}\(\overline{\epsilon}_M \eta + \overline{\epsilon}_M\gamma^{\mu} \eta \gamma_{\mu}\) - (\epsilon \leftrightarrow \eta)
\eea

\vskip0.3cm
%%%%%%%%%%%%%%%%%%%%%%%%%%%%%%%%%%%%%%%%%%%%%%%%%%%%%%%%%%%%%%%%%%%%%%%%%%%%%%%%%%%%%%%%%%%%%%%%%%%%%%%%%%
\section{BLG theory}
The matter content in BLG theory is eight scalar fields $X_I$ and eight fermions $\psi_{\alpha}$ where $I$ transforms as a vector and $\alpha$ as a chiral spinor of the global internal symmetry group SO(8). We denote SO(8) gamma matrices as $\Gamma^I$ and SO(1,2) gamma matrices as $\gamma^{\mu}$. We define the chirality matrix of SO(8) as
\bea
\Gamma &=& \Gamma^{1...8}.
\eea
We denote by $c$ the charge conjugation matrix in SO(1,2). The charge conjugation matrix of SO(8) can be chosen to be unity. The fermions are constrained by
\bea
\Gamma \psi &=& -\psi\cr
\overline{\psi} &=& \psi^T c
\eea
Here $\overline{\psi} = \psi^{\dag}\gamma^0$. If we let $\gamma^0 = c$ this is the SO(8) Majorana-Weyl spinor condition $\psi^{\dag} = \psi^T$, that is all components are real. We let $\epsilon_{\dot{\alpha}}$ denote a supersymmetry parameter,
\bea
\Gamma \epsilon &=& \epsilon\cr
\overline{\epsilon} &=& \epsilon^T c
\eea
which will then also have real components. We let $A_{\mu}$ denote a non-dynamical gauge field and define covariant derivative as $D_{\mu} = \partial_{\mu} + A_{\mu}$. In these conventions the ${\cal{N}}=8$ supersymmetry transformations read
\bea
\delta X_I &=& i\overline{\epsilon}_{\dot{\alpha}}\Gamma_{I\dot{\alpha}\alpha}\psi_{\alpha}\cr
\delta \psi_{\alpha} &=& -\gamma^{\mu} \Gamma_{I\alpha\dot{\alpha}}\epsilon_{\dot{\alpha}}D_{\mu}X_I + \frac{1}{6}\Gamma_{I\alpha\dot{\alpha}}\Gamma_{J\dot{\alpha}\beta}\Gamma_{K\beta\dot{\beta}}\epsilon_{\dot{\beta}}[X_I,X_J,X_K]\cr
\delta A_{\mu} &=& -i \overline{\epsilon}_{\dot{\alpha}}\gamma_{\mu} \Gamma_{I\dot{\alpha}\beta}[\cdot,\psi_{\beta},X_I]
\eea
They close on-shell. In particular the fermionic equation of motion reads
\bea
\gamma^{\mu} D_{\mu} \psi - \frac{1}{2} \Gamma_{IJ} [\psi,X_I,X_J] &=& 0.
\eea

\subsection{Trial BLG theory}\label{Trial}
We can use triality of SO(8) to rotate ${\bf 8}_{\rm v}, {\bf 8}_{\rm s}, {\bf 8}_{\rm c}$. We want to do this in such a way that the ABJM SO(6) R-symmetry is embedded in SO(8) in such a way that we have the decomposition rules
\bea
{\bf 8}_{\rm s} &\rightarrow & {\bf 4}+ {\bf 4}'\cr
{\bf 8}_{\rm c} &\rightarrow & {\bf 4}+{\bf 4}'\cr
{\bf 8}_{\rm v} &\rightarrow & {\bf 6}+{\bf 1}+{\bf 1}
\eea
To this end we make the following triality rotation of matter fields and supersymmetry parameters,
\bea
X_{I} &\rightarrow & X_{\alpha}\cr
\psi_{\alpha} &\rightarrow & \psi_{\dot{\alpha}}\cr
\epsilon_{\dot{\alpha}} &\rightarrow & \epsilon_{I}.
\eea
The BLG theory is then mapped to a trial theory where supersymmetry transformations read
\bea
\delta X_{\alpha a} &=& i\overline{\epsilon}_{I} \Gamma_{I\alpha \dot{\alpha}} \psi_{\dot{\alpha}a}\cr
\delta \psi_{\dot{\alpha}a} &=& -\gamma^{\mu}\Gamma_{I\dot{\alpha}\alpha}\epsilon_I D_{\mu}X_{\alpha a} + \frac{1}{6}\Gamma_{K\dot{\alpha}\alpha}\Gamma_{K\beta \dot{\gamma}}\Gamma_{I\dot{\gamma} \gamma} \epsilon_I [X_{\alpha},X_{\beta},X_{\gamma}]\cr
\delta A_{\mu} &=& -i\overline{\epsilon}_I \gamma_{\mu} \Gamma_{I\alpha \dot{\beta}} [\cdot,\psi_{\dot{\beta}},X_{\alpha}]
\eea
To understand this, one just re-labels indices and defines
\bea
\Gamma_{I\alpha\dot{\alpha}} = \Gamma_{\dot{\alpha}I\alpha} = \Gamma_{\alpha\dot{\alpha}I}\cr
\Gamma_{I\dot{\alpha}\alpha} = \Gamma_{\alpha I \dot{\alpha}} = \Gamma_{\dot{\alpha}\alpha I}.
\eea

To relate to the ABJM supersymmetry transformations, we decompose
\bea
X_{\alpha a} &=& \(\begin{array}{c}
Z^A_a\\
Z_A^a
\end{array}\),\cr
\psi_{\dot{\alpha}a} &=& \(\begin{array}{c}
\psi^{Aa}\\
-\psi_{Aa}
\end{array}\)
\eea
into Weyl spinors of SO(6) and we let
\bea
\epsilon_I &=& (\epsilon^M,\epsilon^7,\epsilon^8).
\eea
A Majorana-Weyl spinor $X$ of SO(8) is subject to
\bea
X^{\dag} &=& X^T.
\eea
We introduce a complex supersymmetry parameter
\bea
\epsilon &\equiv & \epsilon^7+i\epsilon^8
\eea
We can parametrize the six supersymmetries by the supersymmetry parameters
\bea
\epsilon^{AB} &\equiv & \epsilon^M \Sigma^{M,AB},\cr
\epsilon_{AB} &=& \frac{1}{2}\epsilon_{ABCD} \epsilon^{CD}
\eea
These supersymmetry variations become
\bea
\delta Z^A_a &=& -i\overline{\epsilon}^{AB}\psi_{Ba}\cr
\delta Z_A^a &=& i\overline{\epsilon}_{AB}\psi^{Ba}\cr
\delta \psi^{Aa} &=& -\gamma^{\mu}\epsilon^{AB}D_{\mu}Z_{B}^a+\(\epsilon^{AB}Z_B^b Z_C^c Z^C_d + \epsilon^{BC} Z_B^b Z_C^c Z^A_d \) f^{da}{}_{bc}\cr
\delta \psi_{Aa} &=& \gamma^{\mu}\epsilon_{AB}D_{\mu}Z^B_a-\(\epsilon_{AB}Z^B_b Z^C_c Z_C^d +\epsilon_{BC}Z^B_b Z^C_c Z_A^d \) f^{bc}{}_{da}\cr
\delta \t A_{\mu}{}^b{}_a &=& -i\(\overline{\epsilon}^{AB}\gamma_{\mu} \psi_{Ac} Z_B^d-\overline{\epsilon}_{AB}\gamma_{\mu}Z^A_c \psi^{Bd}\)f^{bc}{}_{da}
\eea
We also have two more supersymmetries in trial BLG theory, parametrized by $\epsilon$ and $\epsilon^*$. These are
\bea
\delta Z^A_a &=& \overline{\epsilon}^* \psi^{Aa} \cr
\delta Z_A^a &=& -\overline{\epsilon} \psi_{Aa} \cr
\delta \psi^{Aa} &=& -i\gamma^{\mu}\epsilon^* D_{\mu}Z^A_a+i\epsilon^* Z_B^c Z^{B}_d f^{da}{}_{bc}Z_{A}^b-\frac{i}{3} \epsilon \epsilon^{ABCD} Z_B^b Z_C^c Z_{Dd} f^{da}{}_{bc}\cr
\delta \psi_{Aa} &=& i\gamma^{\mu}\epsilon D_{\mu}Z_A^a-i\epsilon Z^B_c Z_{B}^d f^{bc}{}_{da}Z_{Ab}+\frac{i}{3} \epsilon^* \epsilon_{ABCD} Z^B_b Z^C_c Z^{Dd} f^{bc}{}_{da}\cr
\delta \t A_{\mu}{}^b{}_a &=& -\(\overline{\epsilon} \gamma_{\mu} Z^A_c \psi_A^d + \overline{\epsilon}^* \gamma_{\mu} \psi^A_c Z_A^d\) f^{bc}{}_{da}.
\eea
Now we wrote these BLG supersymmetry variations in an ABJM notation but they are gauge covariant, and close on-shell, only when the structure constants $f^{bc}{}_{da}$ are real and totally antisymmetric, and indices are raised by $\delta^{ab}$.

\end{document}